\documentstyle[12pt,epsf,epsfig,rotate,graphicx]{article}
\def\lsim{\mathrel{\rlap{\lower3pt\hbox{\hskip0pt$\sim$}}
    \raise1pt\hbox{$<$}}}         
\def\gsim{\mathrel{\rlap{\lower4pt\hbox{\hskip1pt$\sim$}}
    \raise1pt\hbox{$>$}}}         
\def\simlt{\mathrel{\raise.3ex\hbox{$<$\kern-.75em\lower1ex\hbox{$\sim$}}}}
\def\simgt{\mathrel{\raise.3ex\hbox{$>$\kern-.75em\lower1ex\hbox{$\sim$}}}}
\newcommand{\be}{\begin{equation}}
\newcommand{\ben}{\begin{subequations}}
\newcommand{\een}{\end{subequations}}
\newcommand{\beq}{\begin{eqalignno}}
\newcommand{\eeq}{\end{eqalignno}}
\newcommand{\ee}{\end{equation}}

\newcommand{\tanb}{\tan \! \beta}
\newcommand{\cotb}{\cot \! \beta}
\newcommand{\mto}{m^2_{\tilde{t}_1}}
\newcommand{\mtt}{m^2_{\tilde{t}_2}}
\newcommand{\mbo}{m^2_{\tilde{b}_1}}
\newcommand{\mbt}{m^2_{\tilde{b}_2}}
\newcommand{\ghat}{\hat{g}^2}
\newcommand{\htop}{h_t^2}
\newcommand{\hb}{h_b^2}
\def\lsim{\:\raisebox{-0.5ex}{$\stackrel{\textstyle<}{\sim}$}\:}
\def\gsim{\:\raisebox{-0.5ex}{$\stackrel{\textstyle>}{\sim}$}\:}
\begin{document}
\begin{titlepage}
\begin{center}
{\Large\bf Phenomenological Issues in Supersymmetry with
Non-holomorphic Soft Breaking}
\vspace{1cm}

$M.~A.~ \c{C}ak{\i}r^{a}$, \,$S.~Mutlu^{a}$ $and$ \, $L.~ Solmaz^{b}$

{\it $^a$ Department of Physics, Izmir Institute of Technology,
IZTECH, Turkey, TR35430}

{\it $^b$ Department of Physics, Bal{\i}kesir University, Bal{\i}kesir, Turkey, TR10100}
\end{center}
\vspace{1cm}
\begin{abstract}
We present a through discussion of motivations for and phenomenological issues in supersymmetric models with
minimal matter content and non-holomorphic soft-breaking terms. Using the unification of the  gauge couplings
and assuming SUSY is broken with non-standard soft terms, we provide semi-analytic solutions of the RGEs for low
and high choices of $tan\beta$ which can be used to study the phenomenology in detail.

We also present a generic form of RGIs in mSUGRA framework which can be used to derive new relations in addition
to those existing in the literature. Our results are mostly presented with respect to the conventional minimal
supersymmetric model for ease of comparison.
\end{abstract}
\vspace {10cm} \small{e-mail:lsolmaz@balikesir.edu.tr}
\end{titlepage}
\section{Introduction}
Supersymmetry is an elegant symmetry for stabilizing the
electroweak scale against strong ultraviolet sensitivity of the
Higgs sector induced by quantum fluctuations. This symmetry, given
that no experiment has yet observed any of the superpartners,
cannot be operative at energies below the Fermi scale. This very
constraint is saturated by breaking global supersymmetry
explicitly via mass parameters ${\cal{O}}({\rm TeV})$ in such a
way that the quadratic divergence of the Higgs sector is not
regenerated. In more explicit terms, the action density of the
minimal supersymmetric model (MSSM) which is based on the
superpotential
\begin{eqnarray}
\label{superpot1} \widehat{W} = h_t\, \widehat{t}_R \widehat{Q}_L
\widehat{H}_u + h_b\, \widehat{b}_R \widehat{Q}_L   \widehat{H}_d
+ h_{\tau}\, \widehat{\tau}_R \widehat{L}_L   \widehat{H}_d + \mu
\widehat{H}_u  \widehat{H}_d
\end{eqnarray}
as obtained after discarding all Yukawa couplings except those of
the heaviest fermions, is augmented by additional terms (see, for
instance, \cite{gordy} for a review)
\begin{eqnarray}
\label{lagran} &&m_{H_u}^2 H_{u}^{\dagger} H_u + m_{H_d}^2
H_{d}^{\dagger} H_d + m_{t_L}^2 \widetilde{Q}_L^{\dagger}
\widetilde{Q}_L + m_{t_R}^2 \widetilde{t}_R^{\dagger}
\widetilde{t}_R + m_{b_R}^2 \widetilde{b}_R^{\dagger}
\widetilde{b}_R +
 m_{\tau_L}^2 \widetilde{L}_L^{\dagger} \widetilde{L}_L + m_{\tau_R}^2
 \widetilde{\tau}_R^{\dagger} \widetilde{\tau}_R+\nonumber\\
&& \left[ h_t A_t \widetilde{t}_R \widetilde{Q}_L {H}_u + h_b A_b
\widetilde{b}_R \widetilde{Q}_L {H}_d + h_{\tau} A_{\tau}
\widetilde{\tau}_R \widetilde{L}_L
 {H}_d + \mu^{\prime} B  {H}_u {H}_d + \sum_{a} \frac{M_a}{2} \lambda_a \lambda_a + \mbox{h.c.} \right]
\end{eqnarray}
which contain massive scalars, gauginos as well as a set of
triscalar couplings among sfermions and Higgs bosons. The
operators in (\ref{lagran}) break supersymmetry in such a way that
Higgs scalar sector does not develop any quadratic sensitivity to
the UV scale.

The soft-breaking terms in (\ref{lagran}) do not necessarily
represent the most general set of operators. Indeed, one may
consider, for instance, triscalar couplings with 'wrong' Higgs as
well as bare Higgsino mass terms. Indeed, such terms have recently
been shown to occur among flux-induced soft terms within
intersecting brane models \cite{camara}. Historically, such terms
have been classified as hard since they have the potential of
regenerating the quadratic divergences \cite{nonholo}. However,
this danger occurs only in theories with pure singlets, and in
theories like the MSSM they are perfectly soft. Hence, the most
general soft-breaking sector must include the operators
\begin{eqnarray}
\mu^{\prime} \widetilde{H}_u \widetilde{H}_d + h_t A_{t^\prime}
\widetilde{t}_R \widetilde{Q}_L {H}_d^{\dagger} + h_b A_{b^\prime}
\widetilde{b}_R \widetilde{Q}_L {H}_u^{\dagger} + h_{\tau}
A_{\tau}^{\prime} \widetilde{\tau}_R \widetilde{L}_L H_u^{\dagger}
+ \mbox{h.c.}
\end{eqnarray}
in addition to those in (\ref{lagran}). Clearly, none of these
operators mimics those contained in the superpotential
(\ref{superpot1}): they are non-holomorphic soft-breaking
operators. Note the structure of the triscalar couplings here; the
triscalar couplings in (\ref{lagran}) are modified by including
the opposite-hypercharge Higgs doublet.

In principle, the theory can contain both $\mu$ and $\mu^{\prime}$
couplings. However, in what follows we will follow the viewpoint
that the $\mu$ parameter is completely soft, that is, $\mu$ in the
superpotential vanishes. This indeed can happen if the theory is
invariant under global chiral symmetries \cite{dimo} at high scale
\cite{gm}. What is crucial about vanishing $\mu$ is that it
automatically solves the $\mu$ problem; the theory does not
contain a supersymmetric mass parameter with a completely unknown
scale. Indeed, in the MSSM stabilization of the $\mu$ parameter to
the electroweak scale requires the introduction of gauge \cite{u1p}- or
non-gauge \cite{leszek} extensions in which the vacuum expectation value (VEV)
of an MSSM gauge-singlet scalar generates an effective $\mu$
parameter. For these reasons, having a nonvanishing $\mu^{\prime}$
in the soft-breaking sector both solves the $\mu$ problem and
serves as if there is a $\mu$ parameter in the superpotential.

The present work is organized as follows. In Appendix A we give the full list of renormalization group equations
(RGEs) for all rigid and soft parameters of the theory (as we hereafter call 'non-holomorphic MSSM' or NHSSM for
short). In Appendix B we list down solutions of the RGEs of all model parameters as a function of their boundary
values taken at the scale of gauge coupling unification $M_{GUT}\approx 10^{16}\, {\rm GeV}$. An important
parameter of the theory is the ratio of the Higgs vacuum expectation values: $\tan\beta \equiv \langle H_u^0
\rangle / \langle H_d^0 \rangle$. In solving the RGEs we will consider low ($\tan\beta = 5$) and high
($\tan\beta = 50$) values of $\tan\beta$ separately. In Sec. 2 we analyze the $Z$ boson mass, in particular, its
sensitivity to GUT-scale parameters. Here we will clarify the differences and similarities between the MSSM and
NHSSM. In Sec. 3 we will discuss sfermion masses in the MSSM and NHSSM for the purpose of identifying their
sensitivities to GUT-scale parameters, in particular, $\mu_0$ and $\mu^{\prime}_0$.
 Neutralinos and charginos are considered in the same section.
In Sec. 4 we will discuss renormalization group invariants in the MSSM and NHSSM in a comparative manner so as
to know what remains scale invariant in two distinct structures. In Sec. 5 we conclude the model.

\section{ Fine-tuning of the $Z$ boson mass: MSSM vs. NHSSM}
It is well known that supersymmetry (SUSY) is not an exact symmerty of Nature, and  there is no unique
mechanism (gravity mediation, gauge mediation, anomaly mediation, etc.) for realizing its breakdown.
From the viewpoint of Non-Standard Soft Breaking in the Minimal Supersymmetric Standard Model (NHSSM),
on one hand, its predictions should  reproduce the SM agreement with data, ensure unification of
gauge couplings at the Grand Unified Theory (GUT) scale with minimal particle content, and on the other,
it should preserve naturalness with soft terms \cite{soft}.

It is expected that in the near future thanks to LHC and its successors, experiments related with superparticle
masses and mixings will yield  enough information  to distinguish between various GUT-models and supersymmetry
breaking mechanisms (see $e.g.$ \cite{LHC}). Taking gravity-mediation as the mechanism responsible for SUSY
breaking, it is important to explore how the soft terms are induced: holomorphic soft terms of the minimal model
or those of the NHSSM with or without $R$ parity violation \cite{drtjones}. In this work we will concentrate on
NHSSM with exact $R$ parity deferring the effects of $R$ parity violation to a future work.

Presently, apart from a number of observables in the flavor-chaning neutral current sector, the $Z$ boson mass is
the main parameter that relates precision measurements to soft masses. In other words, the soft terms must self-organize
so as to reproduce the measured value of the $Z$ boson mass \cite{soft}. Hence, it is profitable to analyze $M_Z$ in
the MSSM and NHSSM in a comparative fashion.

\subsection{Evolution of  soft terms}
For the soft breaking parameters of the NHSSM \cite{soft}, we use one-loop Renormalization  Group Equations
(RGEs) \cite{rge} and thereby express their weak scale values in terms of GUT boundary conditions (see Appendix
\ref{appRGEs}). Once weak scale mass values of SUSY particles are known, it will be possible to make educated
guesses as to the GUT side. Meanwhile, the most general semi-analytic solution set of the RGEs for the NHSSM is
too large for practical purposes to carry out phenomenological analyses which we present in Appendix
\ref{appnhssm}. Nevertheless,  the number of free parameters can be considerably reduced if one assumes the
universality of the soft terms at the GUT scale. In this case solutions are phenomenologically more viable and
they can be found in Appendix \ref{appuni} for all soft terms. Our choice for the GUT scale universality
condition can be stated (dropping the contributions of all fermion generations but the third family) as some
prototype structure inspired from minimal supergravity:
\begin{eqnarray}\label{mintrans}
m_{{H_u},{H_d},{t_L},{t_R},{b_R},{l_L},{l_R}}(0) \rightarrow  m_0\,,\, \mu^{\prime}(0) \rightarrow \mu^{\prime}_0\,,\nonumber \\
A_{t,b,\tau}(0) \rightarrow A_0 \,,\,A_{t,b,\tau}^{\prime}(0) \rightarrow
A_0^{\prime}\,,\,M_{1,2,3}(0)\rightarrow M\,.\end{eqnarray}
Clearly, one may relax all or part of these
conditions whereby obatining a larger parameter space augmenting the results presented in Appendix
\ref{appnhssm}. One should note that even if universal soft masses are assumed at the Planck scale,
consideration of  different boundary conditions for all soft terms including phases is more elegant, but then it
gets difficult to achieve cetain clear-cut statements from the phenomenological side. To evade this cumbersome
reality one needs certain inspirations which can be expected from string models. In order to use the most
general one-loop solutions presented in this work, one can choose  for  instance, if the initial value of
gauginos are not necessarily the same, then $M_{30} \neq M_{20} \neq M_{10}$ can be implemented, and this
approach can be generalized to all soft breaking terms.

One of the most important distinctions is that, in the MSSM none of the soft masses depend on the initial value of $\mu$, whereas in
NHSSM both $A^{\prime}$ parameters
and soft masses do depend on $\mu_0^{\prime}$. Using the universality conditions of (\ref{mintrans}), let us
present some of the soft masses in both of the models for low $tan \beta$ choice ($tan \beta$=5). In the MSSM
masses of up and down Higgs at the weak scale can be expressed using boundary conditions of common gaugino mass,
cubic and soft mass squared terms, \begin{eqnarray} \label{mssmsoft}
&&m_{H_u}^2(t_Z)=-0.087  A_0^2 + 0.38 A_0  M-0.16  m_0^2-2.8 M^2,\,\nonumber\\
&&m_{H_d}^2(t_Z)=-0.0033 A_0^2 + 0.011 A_0 M + 0.99 m_0^2 + 0.49 M^2,\,
 \end{eqnarray}
whereas in the NHSSM also have primed trilinear couplings,
\begin{eqnarray} \label{nhsoft}
&&m_{H_u}^2(t_Z)= -0.087A_0^2+0.1A_0^{\prime 2}-0.16m_0^2-2.8 M^2+0.067A_0^\prime \mu_0^{\prime}+
0.14 \mu_0^{\prime 2}+0.38 A_0 M, \\
&&m_{H_d}^2(t_Z)=-0.0033A_0^2 - 0.37 A_0^{\prime 2} + 0.99 m_0^2 + 0.49 M^2 - 0.31A_0^\prime\mu_0^{\prime} +
0.6 \mu_0^{\prime 2} + 0.011 A_0 M.\,\,\nonumber
\end{eqnarray}
As it is seen in (\ref{mssmsoft},\ref{nhsoft}), at the electroweak scale, the results are the same except
primed trilinear couplings and $\mu_0,\mu_0^{\prime}$ terms. As a matter of fact NHSSM  predictions reduces to
that of MSSM results under the following transformation:
\begin{equation}
\mu^\prime,A_t^\prime,A_b^\prime,A_\tau^\prime \rightarrow \mu\, ,\,\, m_{H_{u,d}}^2\rightarrow m_{H_{u,d}}^2+\mu^2,
\end{equation}
which declares that NHSSM is a beautiful extension of the MSSM. In the NHSSM, notice that the contribution of
$A_0^{\prime 2}$ terms is not of the same order of $A_0^{2}$ terms for all soft masses, hence  trilinear and
primed-trilinear couplings are not symmetric (see Appendix \ref{appuni}). What is more interesting is that, for
both of the models, all soft masses depend heavily on the gaugino masses with the exception of leptons
$m_{l_{L,R}}^2$. Among others $m_{t_L}^2$ is the most sensitive not only for gaugino masses but also for the
initial value of $\mu^\prime$, for the latter $m_{H_d}^2$ is the least sensitive in the NHSSM.

\begin{figure}[htb]
\begin{center}
 \vspace{0.0cm}
    \includegraphics[height=5 cm,width=8cm]{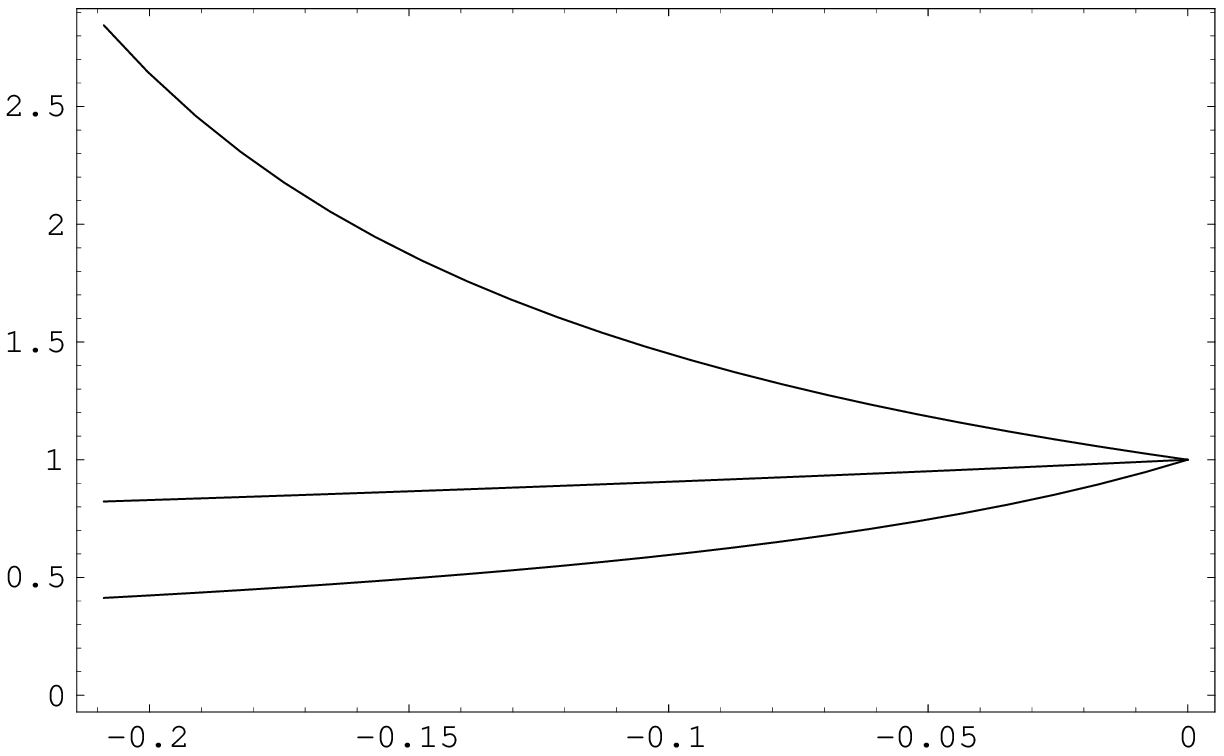}
    \caption[]{Scale dependency of gauginos in both of the models.
     Notice that here the boundary value of M is assumed to be 1 TeV.
      Scale dependency is expressed by dimensionless $t$ such that $t_0$ corresponds to $1.9\times 10^{16}$ GeV .
       Here, Bino is at the bottom, followed by Wino and Gluino.
       Note that the same figure  shows unification of gauge couplings.}
 \begin{picture}(0,0)(0,0)
 \put(-30,72){$t\equiv (4\pi)^{-2}\, \ln({Q}/{Q_0}) $}
 \put(-130,99){\rotate{Gaugino Masses [TeV]}}
  \label{gaufig}
 \end{picture}
    \end{center}
    \vspace{-1.0cm}
\end{figure}
At this point it is appropriate to stress that there are also common model independent predictions like the
evolution of gauiginos (i.e. see Fig.\ref{gaufig}), which stems from the insensitiveness of gauge and Yukawa
RGEs to both of the models at one-loop. On the other hand, trilinear couplings and other soft terms can be seen,
in  a way, to transformed into a new set in which $\mu$ terms are replaced with primed terms.

\subsection{$M_Z$ boundary}

\begin{figure}[htb]
\begin{center}
\vspace{0.0cm}
        \includegraphics[height=4.5 cm,width=7cm]{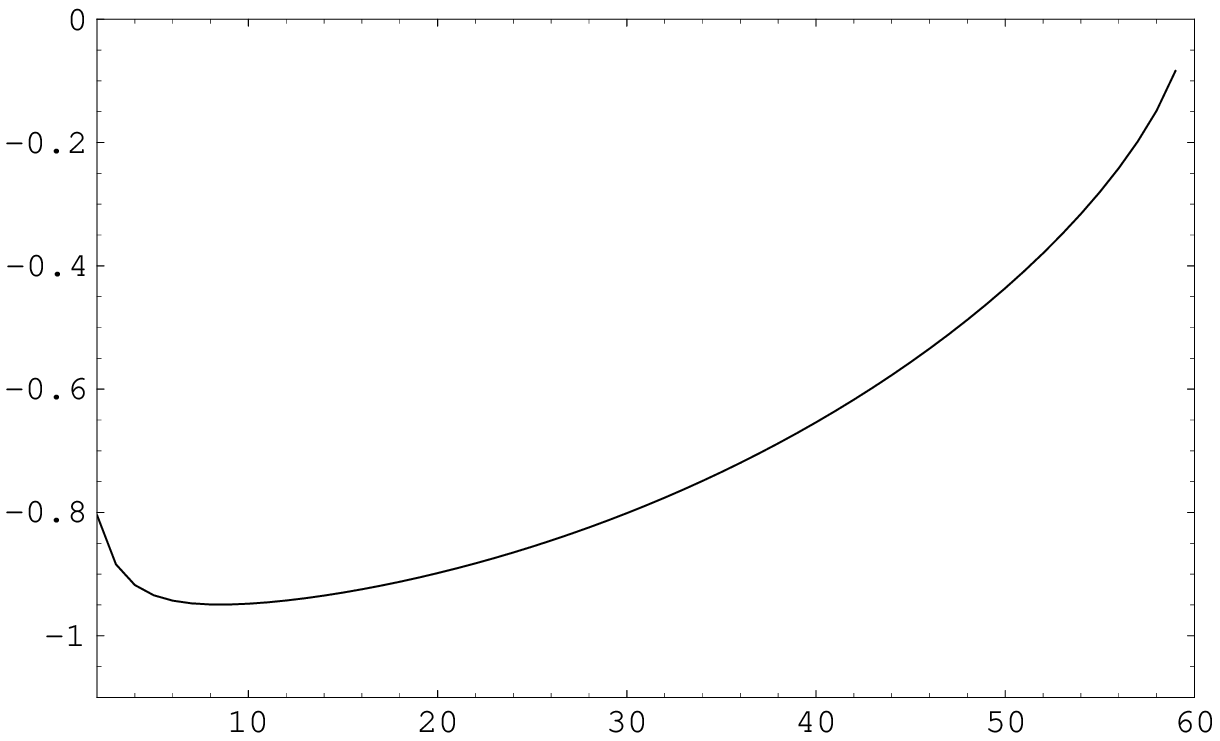}\hspace{0.2cm}
            \includegraphics[height=4.5 cm,width=7cm]{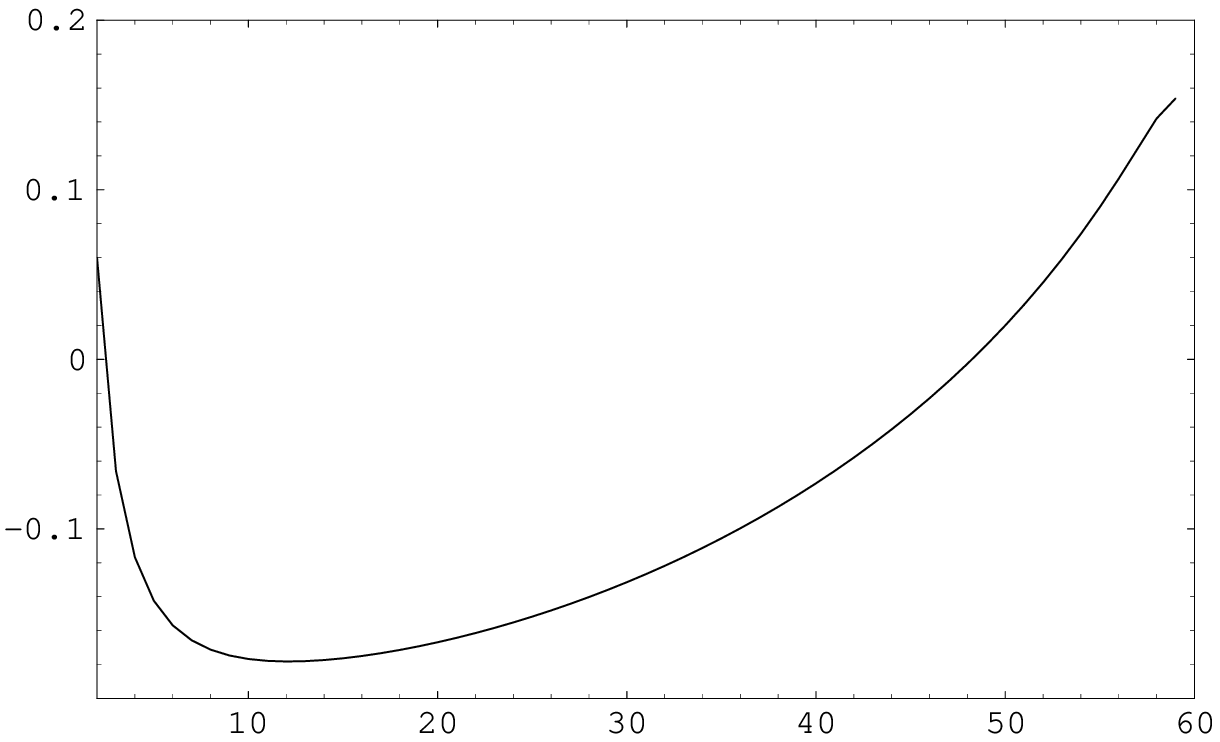}
    \vspace{0cm}
    \caption[]{Evolution of the coefficients of $\mu^2$ terms versus $tan \beta$ $\epsilon$ [2,60] in the MSSM (left),
    and of $\mu^{\prime 2}$ terms in the NHSSM (right) satisfying  $M_Z$ constraint.}
\begin{picture}(0,0)(0,0)
 \put(-120,45){tan$\beta$} 
 \put(-219,115){{$\gamma_6$}} 
  \put(110,45){tan$\beta$} 
 \put(-4,119){{$\gamma_6^\prime$}} 
     \label{figmu}
 \end{picture}
    \end{center}
    \vspace{-1.0cm}
\end{figure}

\begin{figure}[htb]
\begin{center}
\vspace{0.0cm}
    \includegraphics[height=4.5 cm,width=7cm]{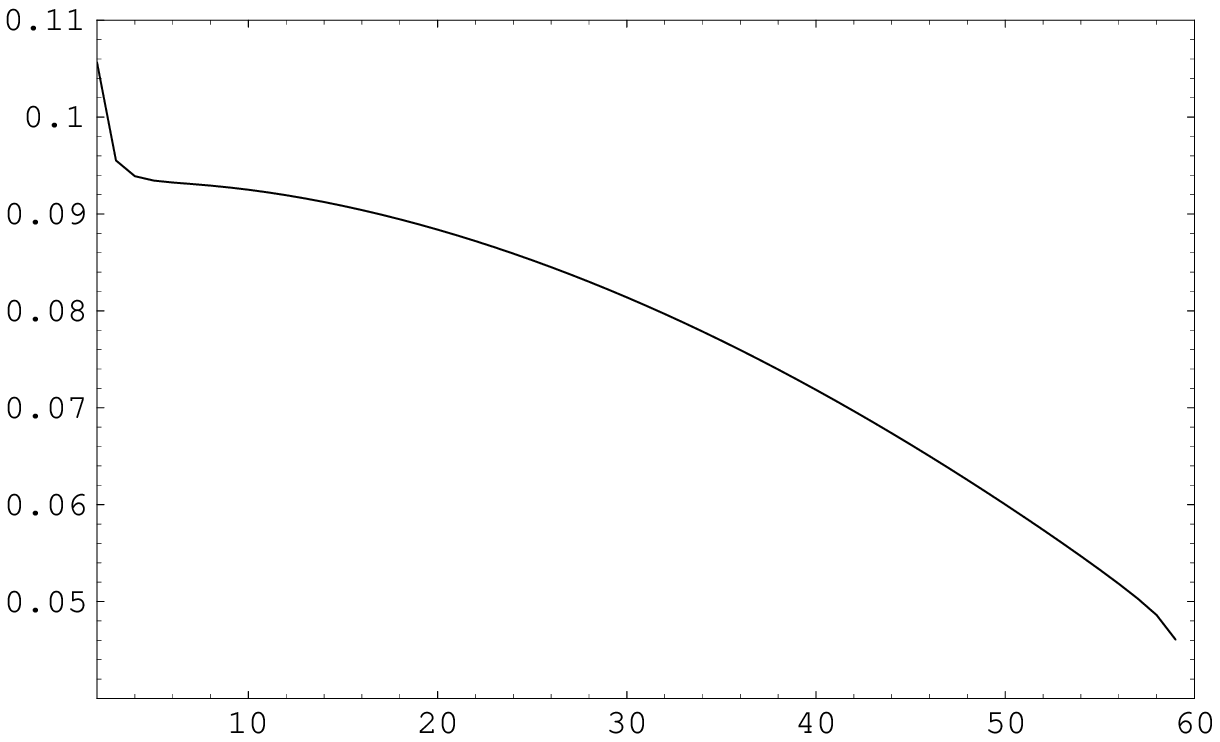}\hspace{0.2cm}
     \includegraphics[height=4.5 cm,width=7cm]{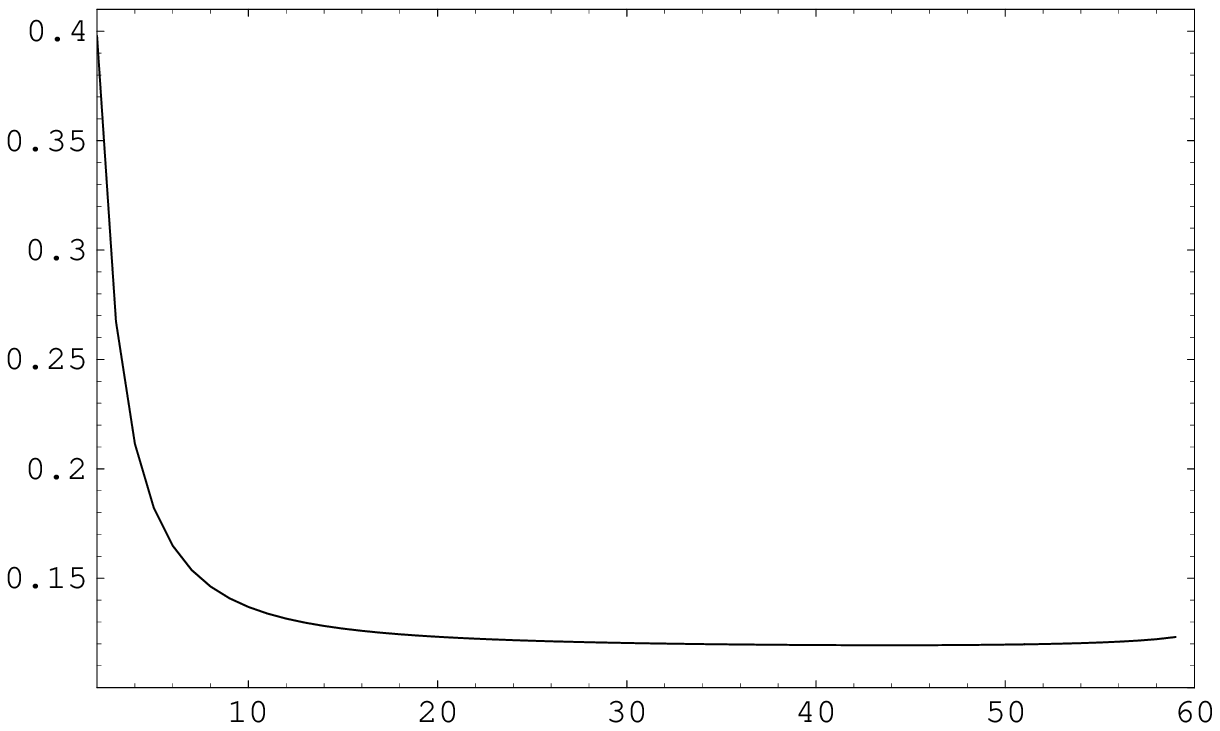}\\\vspace{0.95cm}
     \includegraphics[height=4.5 cm,width=7cm]{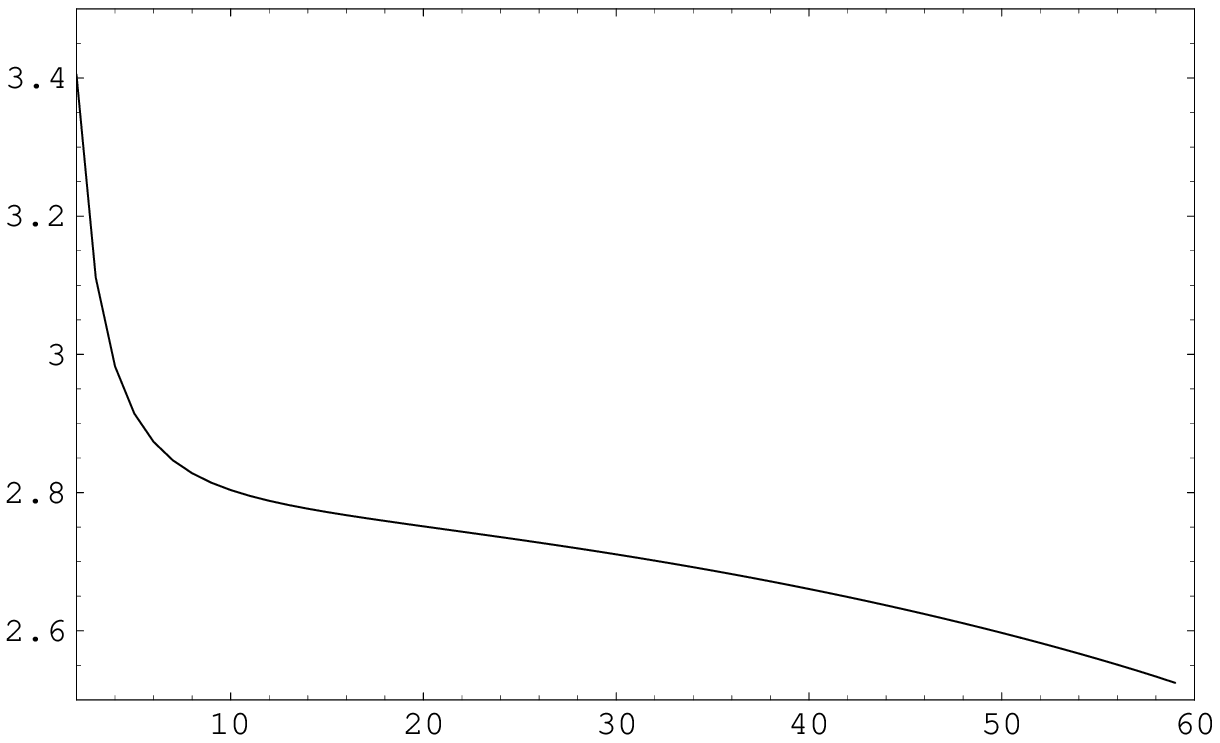}\hspace{0.2cm}
     \includegraphics[height=4.5 cm,width=7cm]{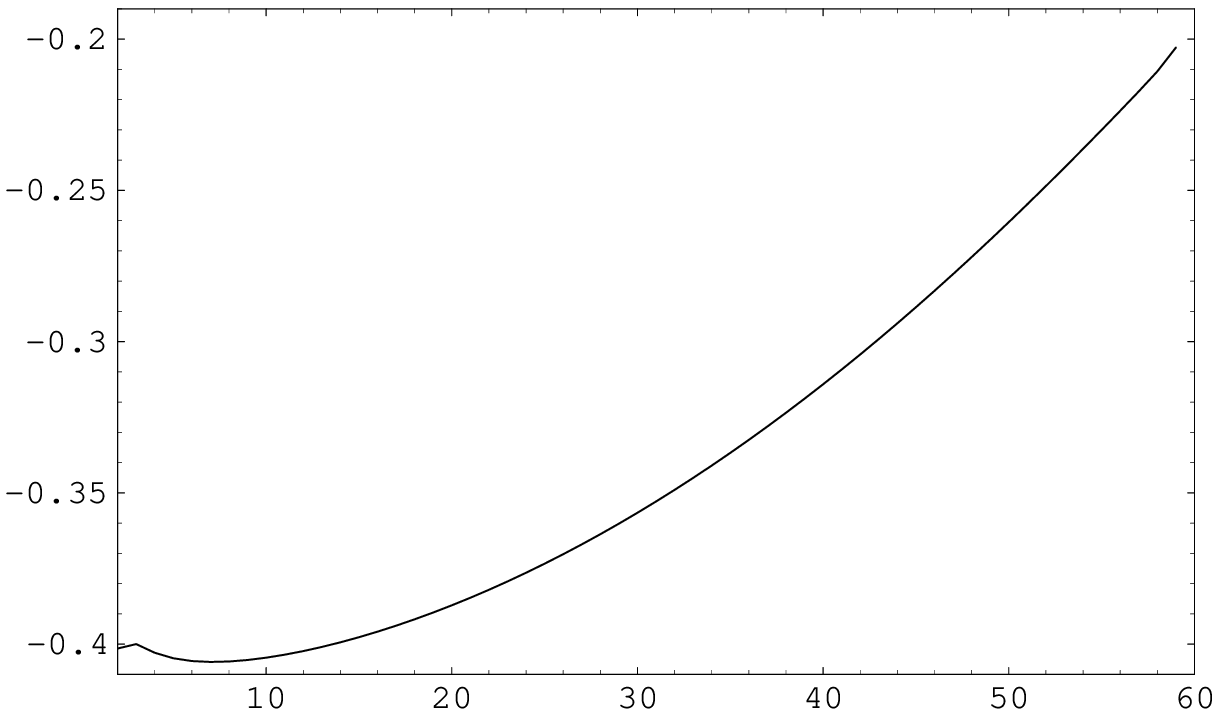}
\vspace{0cm}
    \caption[]{Evolution of the $\gamma_{1,3,4,7}$ terms versus $tan \beta$ $\epsilon$ [2,60]
    in the MSSM}
    \vspace{0.5cm}
\begin{picture}(0,0)(0,0)
 \put(-120,45){tan$\beta$} 
 \put(-219,120){{$\gamma_4$}} 
  \put(110,45){tan$\beta$} 
 \put(-4,120){{$\gamma_7$}} 
  \put(-120,200){tan$\beta$} 
 \put(-219,280){{$\gamma_1$}} 
  \put(110,200){tan$\beta$} 
 \put(-4,280){{$\gamma_3$}} 
   \label{figSM}
 \end{picture}
    \end{center}
    \vspace{-1.5cm}
\end{figure}

\begin{figure}[htb]
\begin{center}
    \includegraphics[height=4.4 cm,width=7cm]{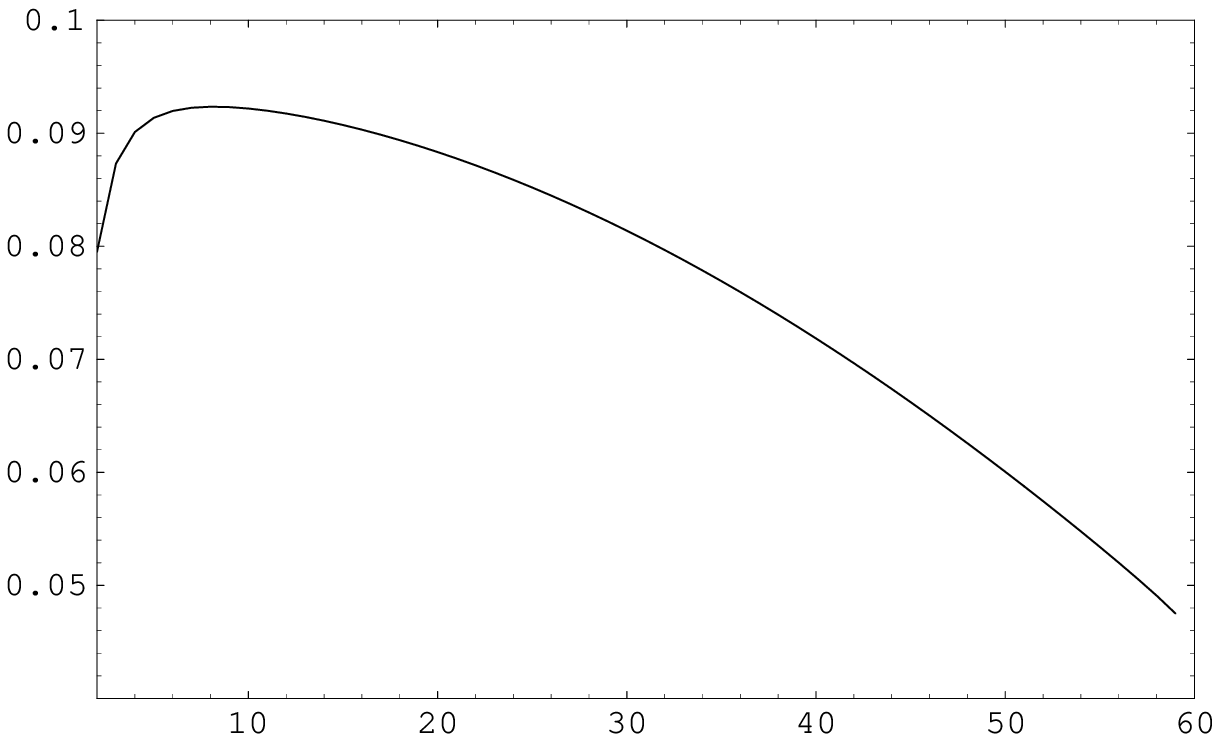}\hspace{0.2cm}
       \includegraphics[height=4.4 cm,width=7cm]{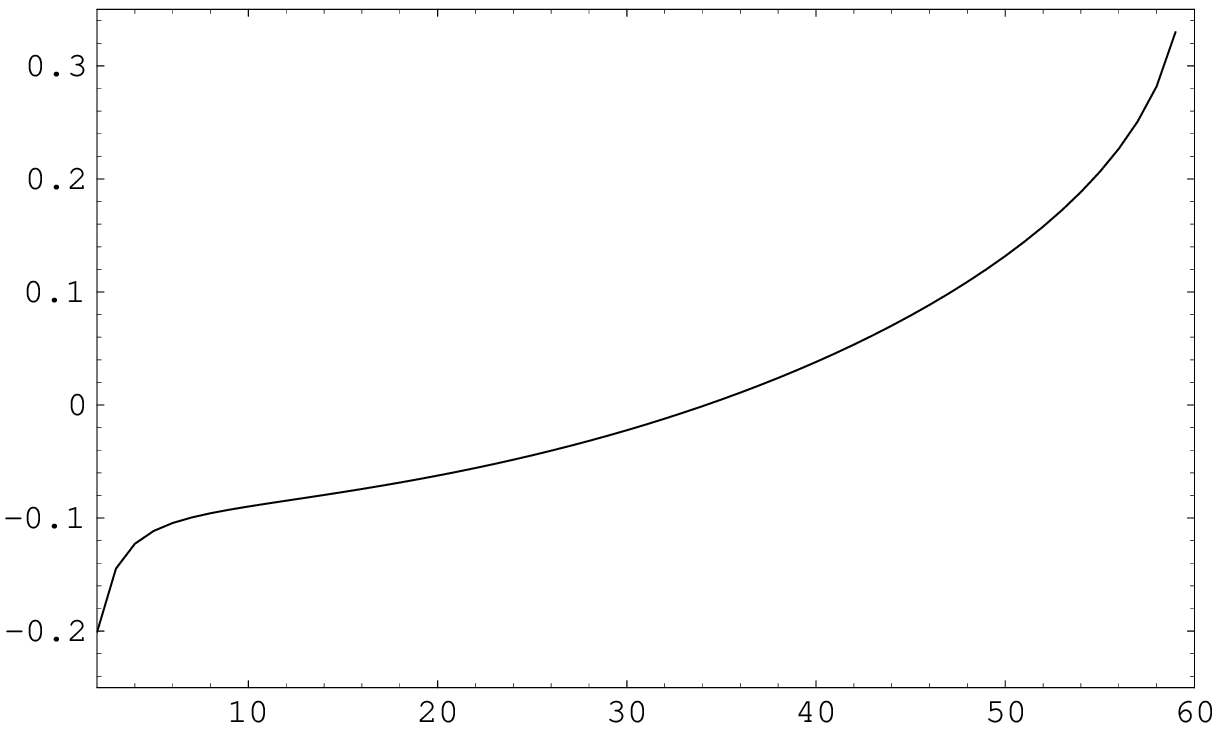}\\\vspace{0.6cm}
       \includegraphics[height=4.4 cm,width=7cm]{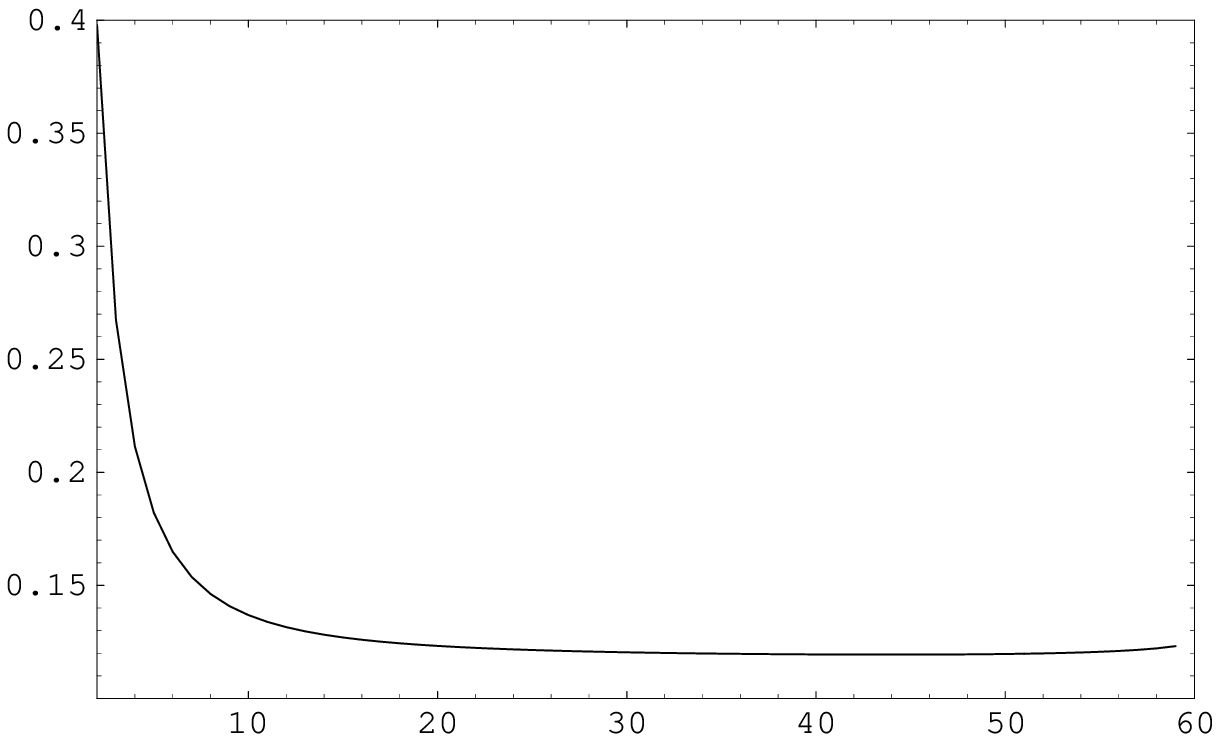}\hspace{0.2cm}
        \includegraphics[height=4.4 cm,width=7cm]{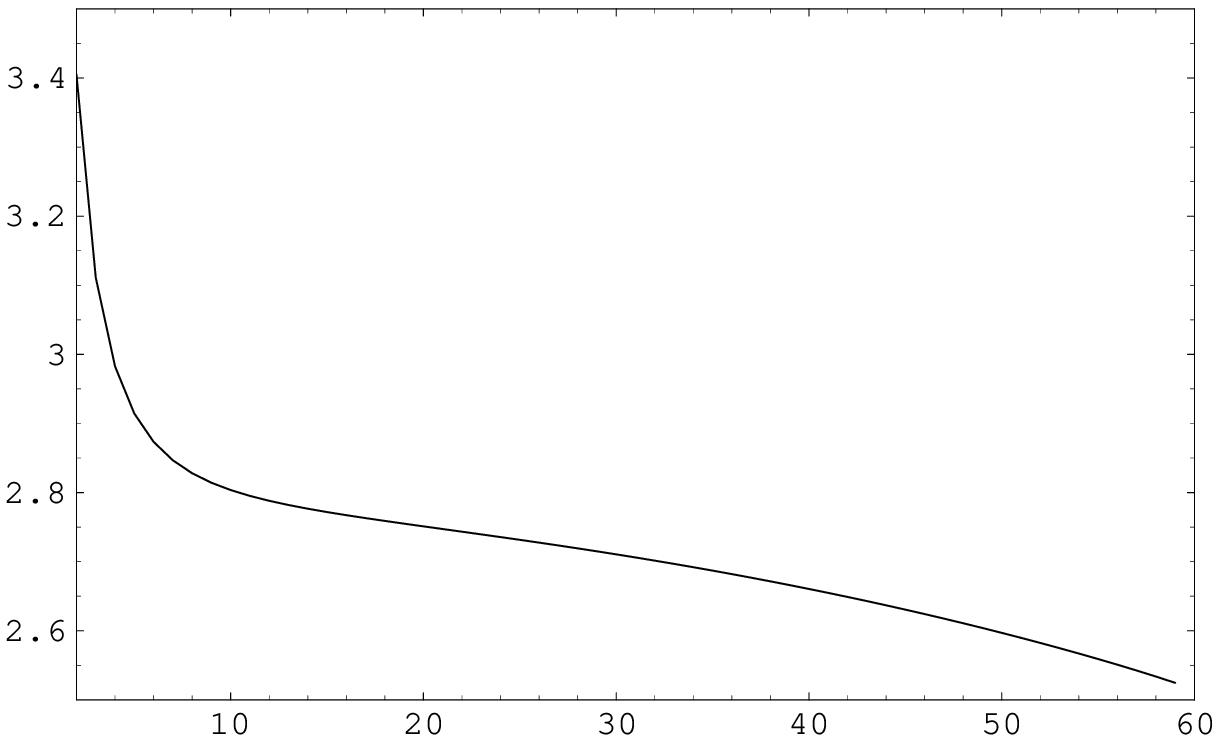}\\\vspace{0.6cm}
        \includegraphics[height=4.4 cm,width=7cm]{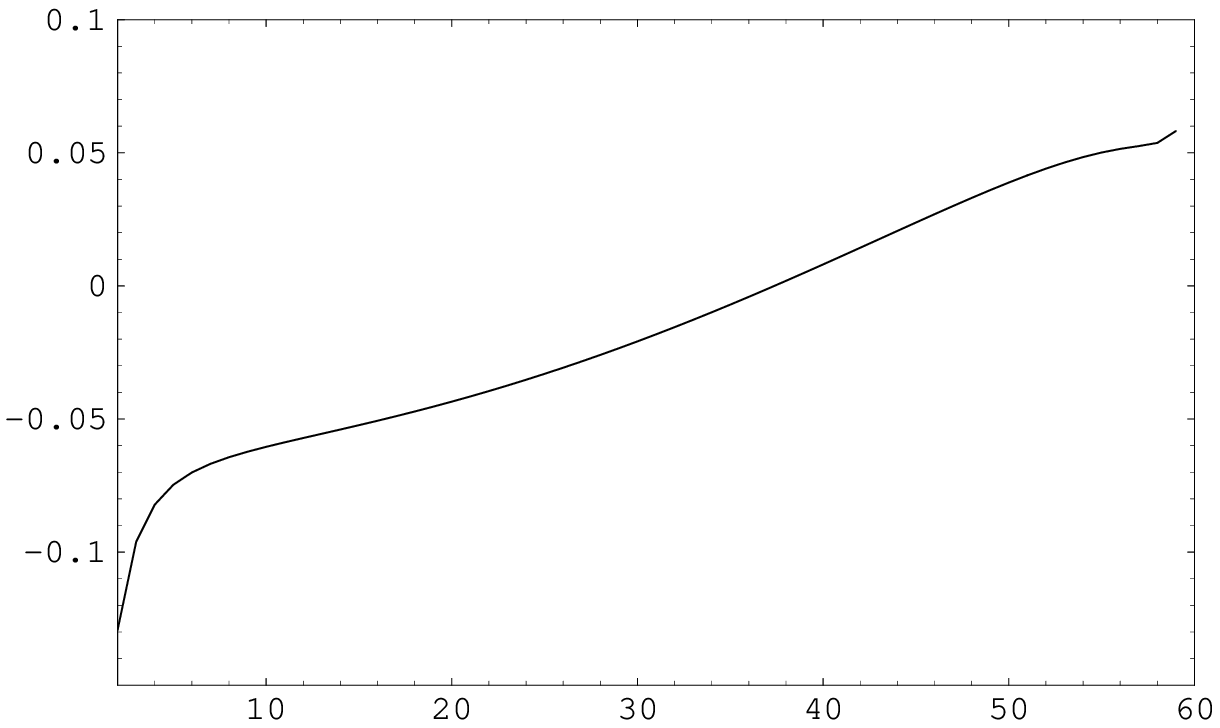}\hspace{0.2cm}
           \includegraphics[height=4.4 cm,width=7cm]{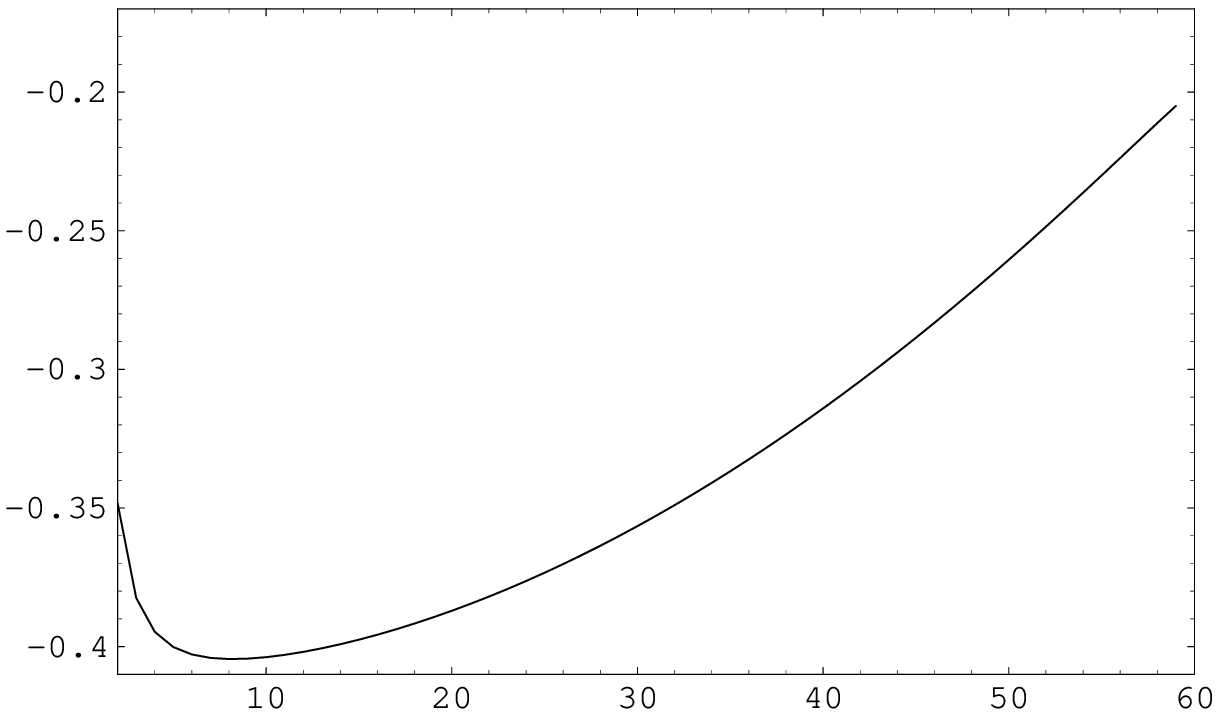}
    \caption[]{Evolution of the $\gamma^\prime_{1,2,3,4,5,7}$ terms versus $tan \beta$ $\epsilon$ [2,60]
    in the NHSSM}
\begin{picture}(0,0)(0,0)
 \put(-120,30){tan$\beta$} 
 \put(-219,105){{$\gamma_5^\prime$}} 
  \put(110,30){tan$\beta$} 
 \put(-4,105){{$\gamma_7^\prime$}} 
  \put(-120,172){tan$\beta$} 
 \put(-219,250){{$\gamma_3^\prime$}} 
  \put(110,172){tan$\beta$} 
 \put(-4,250){{$\gamma_4^\prime$}} 
  \put(-120,315){tan$\beta$} 
 \put(-219,390){{$\gamma_1^\prime$}} 
 \put(110,315){tan$\beta$} 
 \put(-4,390){{$\gamma_2^\prime$}} 
     \label{figNH}
 \end{picture}
    \end{center}
    \vspace{-1.3cm}
\end{figure}

For both of the models, as one of the most crucial constraints for the SM agreement with data, mass of the Z
boson should be considered first, for a successful electroweak symmetry breaking.  Notice that in the MSSM, in
order to get the observed value of $M_Z$, a delicate cancellation between the Higgs masses and $\mu$ is
required, which is the famous $\mu$ problem (see i.e. \cite{muproblem},\cite{gm}). Instead of $\mu$ parameter of
the MSSM, NHSSM bears $A_{{t^\prime}},A_{b^{\prime}},A_{\tau^{\prime}}$ and $\mu^{\prime}$ and its interesting
effect can be seen by minimizing the scalar potential of the NHSSM which brings the constraint
\begin{eqnarray} \frac{M_Z^2(t_Z)}{2}=\frac{ m_{H_d}^2(t_Z)-tan^2\beta\, m_{H_u}^2(t_Z)}{tan^2\beta-1}.\, \end{eqnarray}
The $Z$ boson mass depends on $\mu_0$ rather strongly in the MSSM. As an example for $tan \beta$=5, MSSM
constraints can be expressed under the assumption of universality as
\begin{eqnarray} \frac{M_Z^2(t_Z)}{2}=0.09 A_0^2 + 0.21 m_0^2 + 3 M^2 - 0.92 \mu_0^{2} - 0.39 A_0 M. \end{eqnarray}
However, in NHSSM it does depend on $\mu^{\prime}$ rather weakly $e.g.$ a 10$\%$ change in $\mu_0^{\prime 2}$
generates only a 0.1$\%$ shift in $M_Z^2/2$. To make a comparison, in the NHSSM for the same value of $tan
\beta$:
\begin{eqnarray} \frac{M_Z^2(t_Z)}{2}=0.09 A_0^2 - 0.12 A_0^{\prime 2} + 0.21 m_0^2 + 3 M^2 -
  0.082 A_0^\prime \mu_0^{\prime} - 0.12 \mu_0^{\prime 2} - 0.39 A_0 M. \end{eqnarray}
For the sake of visualization of the NHSSM and MSSM reactions we define dimensionless quantities $\gamma_i(tan\beta)$ such
that the Z constrain can be expressed as
\begin{eqnarray} \frac{M_Z^2(t_Z)}{2}=\gamma_1^\prime A_0^2 + \gamma_2^\prime A_0^{\prime 2} + \gamma_3^\prime m_0^2 +
\gamma_4^\prime M^2 +\gamma_5^\prime A_0^\prime \mu_0^{\prime} + \gamma_6^\prime \mu_0^{\prime 2}
+\gamma_7^\prime A_0 M, \end{eqnarray}
which can be used also for MSSM with obvious modifications. In the range $tan \beta$ $\epsilon$ [2,60], weights of  $\gamma$'s
can be inferred from Figs.\ref{figmu},\ref{figSM} and \ref{figNH}.

In addition  to relaxing sensitivity on the $\mu_0$ terms, we observe that  $tan \beta$ changes the sign of the
$\mu^\prime$ contribution in the NHSSM, and this situation  has important consequences on the model building business.
Note that in the MSSM contribution of $\mu^2$ terms is always destructive (assuming it is real), whereas by
staring the oscillatory behaviour of $\mu^{\prime 2}$ with different choices of $tan \beta$ (see
Fig.\ref{figmu}) one can find a specific prediction for $tan \beta$ such that $\mu^{\prime 2}$ dependency of the
$M_Z^2$ completely vanishes in low and high regions, in addition to destructive or constructive contribution
regions. Such special points can be called as turning points and this corresponds to $\sim49.25$ for high
$tan\beta$ in the NHSSM under the assumption of universal terms. Of course relaxing the universality assumption
brings different turning points.

In addition to capability of getting
rid off $\mu^{\prime}$ terms for specific angles, NHSSM deserves new phenomenological approach which can be
inferred from the Figs.\ref{figmu},\ref{figSM} and \ref{figNH}. In the figures $tan\beta$ evolution of the coefficients of mass dimension 2
terms satisfying the $M_Z$ constraint is given. Behavior of other terms should also be taken into account, which
can be performed using the Fig.\ref{figNH} to unfold the reaction of the model satisfying the mass boundary of the
Z boson. To make a comparison with the MSSM we also presented the $tan\beta$ evolution of similar coefficients in
Fig.\ref{figSM}. Using figures presented here, one can find appropriate regions that satisfy the Z constrain
in both of the models for any $tan \beta$ in the range [2,60]. Nevertheless, one should notice that
constraints presented here are at the tree level and in the universal region, which might change when
 radiative corrections or anomaly  boundary conditions are considered.

Consequently, supersymmetry breaking with non-standard soft terms  has an important virtue of reducing the
sensitivity of $M_Z^2$ to the initial value of the $\mu$ parameter. However, in both cases, the MSSM and NHSSM,
the $Z$ boson mass exhibits a strong sensitivity of the gaugino masses. This follows mainly from the asymptotic
freedom of color gauge group.

\section{Spectrum of sparticles in Minimal Supergravity:\\
MSSM vs. NHSSM}

From the viewpoint of realistic model building  approach any  model should  satisfy other collider bounds besides $M_Z$,
 however we know from direct searches that no supersymmetric particle is observed yet, which can not
set tight bounds on the spectrum of masses of SUSY particles \cite{spect}. Meanwhile mass of Higgs boson can be considered as  on the verge of
experimental verification if  low scale supersymmetry really exists.
We consider particle data group restrictions on the mass of sparticles and simply accept the lower bounds of LEP 2
$m_{soft}>100$ GeV, for the lightest chargino and neutralino half of Z boson width is accepted \cite{PDG}.
 For simplicity and clarity, again, in this section we require all scalars to acquire a common mass
$m_0$, all gauginos to be mass-degerate with $M$, all triscalar couplings to be $A_0$ and all non-holomorphic
triscalars to be $A_0^{\prime}$ all fixed at the GUT scale. In fact,  suppression of the flavor-changing neutral currents as well as
the absence of permanent electric dipole moments already imply that
the soft-breaking masses cannot be all independent and arbitrarily
distributed; they must be correlated by some organizing principle
operating at the unification scale or above. With this assumption one can predict mass of
lightest Higgs boson at tree level using the scalar Higgs potential of the NHSSM which brings the constraints
\begin{eqnarray}
m_{H_d}^2=m_3^2 \,tan\beta-(M_Z^2/2)\, cos2\beta, \\
m_{H_u}^2=m_3^2 \,cot\beta+(M_Z^2/2)\, cos2\beta.
\end{eqnarray}
 During the numerical investigation, we look for real and positive
soft terms in the range $[0,1000]$ GeV,  which results in successful electroweak symmetry breaking patterns for
low $tan\beta$ option. In this case by noting the collider lower bounds on the mass spectrum, parameter space
can be restricted to a good extend, without additional assumptions (like no-scale \cite{noscale}, or some other string inspired
models). With the same range proposed for GUT boundaries there is no succesfull candidate in high $tan\beta$
region, while the universality assumption of (\ref{mintrans}) in charge. When the electroweak symmetry is broken
mass eigenstate of the lightest neutral scalar should satisfy $m_h^0>$ 114 GeV with radiative corrections. By
expanding the scalar potential around the minimum tree-level masses of the fields can be found as
\begin{equation}
m_{A^0}^2=2 m_3^2/ sin 2\beta,
\end{equation}
\begin{equation}
m_{H^\pm}^2=m_{A^0}^2+M_W^2,
\end{equation}
\begin{equation}\label{higgs0}
m_{h^0,H^0}^2=\frac{1}{2}(m_{A^0}^2+m_{Z}^2 \mp \sqrt{(m_{A^0}^2+m_{Z}^2)^2-4 M_Z^2 m_{A^0}^2 cos^2 2\beta}),
\end{equation}
when one-loop quantum corrections are considered SM like Higgs boson gets the largest contributions from t and b squarks.
Notice that without quantum corrections mass of the lightest Higgs boson can not satisfy the experimental boundary, hence
 we study this issue in section \ref{dethiggs} for NHSSM without CP violation; MSSM results including CP violation can be found in \cite{durmush,kore}
Analytic forms of $m_{\tilde{t}_1}$ and $ m_{\tilde{t}_2}$ is given in the following subsection which will be needed in correction business.
\subsection{Sfermions}
For scalar fermions the relation between gauge eigenvalues and mass eigenvalues of the NHSSM particles can be
read from the mass-squared matrices. Following that aim, we  provide explicit expressions for the
mass-squared matrices of squark and sleptons using reference \cite{drtjones}. The stop matrix is:
\begin{equation}
\pmatrix{ m_{t_L}^2+ m_t^2 + {1\over6}(4M_W^2 -M_Z^2)\cos2\beta&m_t(A_t-A_{t^\prime} \cot\beta )\cr m_t ( A_t -
A_{t^\prime} \cot\beta ) & m_{t_R}^2 + m_t^2 - {2\over3}(M_W^2 - M_Z^2)\cos 2\beta\cr}.\end{equation}  for which
we obtain the following eigenvalues
\begin{eqnarray}\label{e4a}
m_{\tilde{t}_{1,2}}^2&=&\frac{1}{12} \{6 (2 m_t^2+m_{tL}^2+m_{tR}^2)+3 M_Z^2 cos 2\beta \\
&\mp& \sqrt{\sigma_1 cos 2\beta \left(12 \sigma_2+\sigma_1 cos 2\beta\right) + 36\left[4 A_t^2 m_t^2+\sigma_2^2+4 A_{t^\prime }m_t^2 cot\beta \left(-2 A_t+A_t^\prime cot\beta\right)\right]}\},\,\nonumber
\end{eqnarray}
where $\sigma_1=8 M_W^2-5 M_Z^2$ and $\sigma_2=m_{tL}^2-m_{tR}^2$. Similarly for the bottom squarks we have:
\begin{equation}
\pmatrix{ m_{t_L}^2 + m_b^2 - {1\over6}(2M_W^2 + M_Z^2)\cos 2\beta & m_b ( A_b - A_{b^\prime} \tan\beta )\cr m_b
( A_b - A_{b^\prime} \tan\beta ) & m_{b_R}^2 + m_b^2 + {1\over3}(M_W^2 - M_Z^2)\cos 2\beta\cr}\end{equation}
with eigenvalues
\begin{eqnarray}\label{e4b}
m_{\tilde{b}_{1,2}}^2&=&\frac{1}{12} \{6 (2 m_b^2+m_{tL}^2+m_{bR}^2)-3 M_Z^2 cos 2\beta  \\
&&\mp \sqrt{\sigma_3 cos 2\beta (12\sigma_4-\sigma_3 cos 2\beta) + 36\left[4 A_b^2 m_b^2+\sigma_4^2+4 A_{b^\prime }m_b^2 tan\beta (-2 A_b+A_b^\prime tan\beta)\right]}\},\,\nonumber
\end{eqnarray}
where $\sigma_3=4 M_W^2 - M_Z^2$ and $\sigma_4=m_{bR}^2-m_{tL}^2$. For the tau sleptons we have:
\begin{equation}
\pmatrix{ m_{l_L}^2 + m_{\tau}^2 - {1\over2}(2M_W^2 - M_Z^2)\cos 2\beta & m_{\tau} ( A_{\tau} -
A_{\tau}^{\prime} \tan\beta )\cr m_{\tau} ( A_{\tau} - A_{\tau}^{\prime} \tan\beta ) &m_{l_R}^2 + m_{\tau}^2 +
(M_W^2 - M_Z^2)\cos 2\beta\cr}.\end{equation}
for which eigenvalues can be written as
\begin{eqnarray}
m_{\tilde{\tau}_{1,2}}^2&=&\frac{1}{4} \{2 (2 m_b^2+m_{lL}^2+m_{lR}^2)- M_Z^2 cos 2\beta  \\
&&\mp \sqrt{\sigma_5 cos 2\beta (\sigma_5-4 \sigma_6 cos 2\beta) + 4\left[4 A_\tau^2 m_\tau^2+\sigma_6^2+4
A_{\tau}^\prime m_\tau^2 tan\beta (-2 A_\tau+A_\tau^\prime tan\beta)\right]}\},\,\nonumber
\end{eqnarray}
where $\sigma_5=4 M_W^2 -3 M_Z^2$ and $\sigma_6=m_{lL}^2-m_{lR}^2$.
Explicit expressions related with each of the  elements of these matrices can be extracted from the Appendix
\ref{appuni} of this work for low and high $tan \beta$ choices. In the MSSM sfermion masses depend on
$\mu_0$ only via their (1,2) and (2,1) entires whereas in the NHSSM $\mu_0^\prime$ appears in all entires including
(1,1) and (2,2). When all the Yukawa couplings are set to zero, except $h_t$ and $h_\tau$, it is interesting to
observe SUSY loop effects on the mass squared terms (see \cite{Ref07} and \cite{Ref08}).
\subsection{Charginos and Neutralinos}
The last step is to compare the mass eigenvalues of neutralinos and charginos.
Neutralino values can be read from the following matrix, which resembles the mixing of Higgsinos and neutral gauginos
\begin{equation}
\pmatrix{M_1&0&-M_Z \,cos\beta \, sin\theta_W & M_Z\, sin\beta\,sin\theta_W \cr 0 & M_2 & M_Z\, cos\beta\, cos\theta_W &
-M_Z \,sin\beta\, cos\theta_W \cr -M_Z\, cos\beta \, sin\theta_W & M_Z\, cos\beta\,cos\theta_W & 0& -\mu^\prime \cr M_Z\,
sin\beta\,sin\theta_W & -M_Z\, sin\beta \, cos\theta_W &-\mu^{\prime} & 0\cr}.\end{equation}
Similarly charginos are mixtures of charged Higgsinos and charged gauginos with the mass matrix
\begin{equation}\pmatrix{M_2& \sqrt{2} M_W sin\beta \cr
\sqrt{2} M_W sin\beta& \mu^{\prime}\cr}.\end{equation}
Since we assume R-parity conservation LSP is the lightest neutralino. Explicit form of matrix elements
can be found in Appendices for low and high values of $tan \beta$.

\subsection{Higgs boson mass and LEP bounds}
\label{dethiggs}
In this section we will compute the Higgs boson mass in
NHMSSM. The main impact of the non-holomorphic soft terms
on the Higgs boson masses stems from the modifications
in the sfermion mass matrices. Indeed, as one infers
from the forms of the sfermion mass-squared
matrices in Sec. 3.2, the mixing between the left and right-handed
sfermions are described by the holomorphic triscalar coupling
$A_t$ and the non-holomorphic contribution $A_f^{\prime}$. The
left-right mixing thus changes from flavor to flavor in contrast
to MSSM where $A_f^{\prime}$ is replaced by flavor-insensitive
quantity $\mu$ parameter.

For a proper understanding of the Higgs sector it is necessary to implement the loop corrections as otherwise
the tree level masses turn out to be too low to saturate the experimental bounds. The radiative corrections to
Higgs boson masses and couplings have already been computed in \cite{durmush,kore} including the CP-violating
effects. Concerning the neutral Higgs sector, it is useful to use the parametrization \be \label{e1} H_d^0 =
\frac{1} {\sqrt{2}} \left( \phi_1 + i \varphi_1 \right); \,\, H_u^0 = \frac {1} {\sqrt{2}} \left( \phi_2 + i
\varphi_2 \right), \ee where $\phi_{1,2}$ and $\varphi_{1,2}$ are real fields. The Higgs potential, including
the Coleman-Weinberg contribution, reads as
\begin{eqnarray} \label{e2} V_{\rm Higgs} &=& \frac{1}{2} m_{H_d}^2 |H_{d}^0|^2 + \frac{1}{2}
m_{H_u}^2 |H_{u}^0|^2 - (m_{3}^2 H_{u}^0 H_{d}^0+ c.c.)
  \nonumber \\
&&+ \frac {g^2 + g'^2} {8}\left( |H_{d}^0|^2-|H_{u}^0|^2\right)^2 + \frac {1} {64 \pi^2} {\rm Str} \left[ {\cal M}^4 \left( \log \frac { {\cal M}^2} {Q_0^2} -
\frac{3}{2} \right) \right], \end{eqnarray}
where $g$ and $g'$ stand for the $SU(2)$ and $U(1)_Y$ gauge couplings, respectively (${g'}^2 = \frac{3}{5}
g_1^2$). $Q_0$ in (\ref{e2}) is the renormalization scale, and ${\cal M}$ is the field--dependent mass matrix of
all modes that couple to the Higgs bosons. The masses of the quarks are to be taken into consideration  of which
the most  important contributions come from:
\be \label{e3} m_b^2 = \frac{1}{2} \hb \left( \phi_1^2 + \varphi_1^2 \right); \ \ \ \ m_t^2 = \frac{1}{2} \htop
\left( \phi_2^2 + \varphi_2^2 \right). \ee Now, using the eigenvalues of the field-dependent squark mass
matrices (\ref{e4a},\ref{e4b}) in (\ref{e2}) one can systematically compute the Higgs boson masses at the
minimum of the potential obtained via the conditions
 \begin{eqnarray}
\frac{ \partial V_{Higgs}}{\partial \phi_1 }=0,\,\frac{\partial V_{Higgs}}{\partial \phi_2}=0
 \end{eqnarray}
with  $\langle \varphi_1 \rangle = \langle \varphi_2 \rangle = 0$ and
\be \label{e6nn} \langle \phi_1 \rangle^2 + \langle \phi_2 \rangle^2 = \frac {M_Z^2} {\ghat} \simeq (246 \ {\rm
GeV})^2, \,\, \frac{ \langle \phi_2 \rangle} {\langle \phi_1 \rangle} = \tanb, \ee where $\ghat=(g^2+g^{\prime
2})/4$. The mass matrix of the neutral Higgs bosons are computed from the matrix of second derivatives of the
potential (\ref{e2}). Notice that after including the one-loop corrections to the Higgs potential, the Z mass
becomes dependent on the top- and stop quark masses too \cite{boer}. In this case there will be a correction
term
\begin{eqnarray} \frac{M_Z^2(t_Z)}{2}=\frac{ m_{H_d}^2(t_Z)-tan^2\beta\, m_{H_u}^2(t_Z)-\Delta_Z^2(t,b)}{tan^2\beta-1}.\, \end{eqnarray}
 where
 \begin{eqnarray}
 \Delta_Z^2(t)=\frac {3 g^2 m_t^2} {32 \pi^2 M_W^2}\left[\left(A_t^2-A_{t^\prime}^2 cot^2\beta \right)\frac {f(m^2_{\tilde{t}_1})-f(m^2_{\tilde{t}_2})} {m^2_{\tilde{t}_1}-m^2_{\tilde{t}_2}}+2 m_t^2+f(m^2_{\tilde{t}_1})+f(m^2_{\tilde{t}_2})\right]
 \end{eqnarray}
 and
\be \label{e8} f(m^2) = 2 m^2 \left( \log \frac {m^2} {Q_0^2} - 1 \right). \ee
Similarly $\Delta_Z^2(b)$ can be found with the $t\rightarrow b$ substitution. This corrections require a large amount
of fine tuning if the mass splitting between the particles and sparticles is large \cite{soft}.

The Goldstone boson $G^0 = \varphi_1 \cos \beta - \varphi_2 \sin \beta$ is swallowed by the $Z$ boson. We are
then left with a squared mass matrix ${\cal M}_H^2$ for the three states $\varphi = \varphi_1 \sin \beta +
\varphi_2 \cos \beta, \ \phi_1$ and $\phi_2$. If the theory has CP-violating phases (via the phases of the
triscalar couplings and $\mu^{\prime}$) the $\varphi$ mixes with $\phi_{1}$ and $\phi_2$. In the CP-conserving
limit, however, $\varphi$ decouples from the rest, and assumes the mass-squared:
\be \label{e10} \left. {\cal M}^2_{H} \right|_{aa} = m_A^2 = \frac {2 m_{3}^2} {\sin(2\beta)} + \frac {2} {\sin(2 \beta)}
\left[ \htop A_t  A_{t^\prime } F(\mto,\mtt) + \hb A_b  A_{b^\prime } F(\mbo,\mbt) \right], \ee
where
\be \label{en1} F(m_1^2,m_2^2) = \frac {3} {32 \pi^2} \frac {f(m_1^2)-f(m_2^2)} {m_2^2-m_1^2}. \ee

The remaining real scalars $\phi_1$ and $\phi_2$ mix with each other via the mass-sqaured matrix:
\begin{eqnarray}
\label{e15}
\left. {\cal M}^2_H \right|_{\phi_1 \phi_1} &=& M_Z^2 \cos^2 \beta + m_A^2 \sin^2 \beta \nonumber \\
&+& \frac {3 m_t^2} {8 \pi^2} \left[ g(\mto,\mtt) R_t \left( \htop R_t -\cotb X_t \right) + \ghat \cotb R_t \log \frac {\mtt} {\mto} \right]\nonumber \\
&+& \frac {3 m_b^2} {8 \pi^2} \left\{ \hb \log \frac {\mbo \mbt} {m_b^4}- \ghat \log \frac {\mbo \mbt} {Q_0^4}\right. \label{e15a} \\
&& \left.
+ g(\mbo,\mbt) R_b' \left( \hb R_b' + X_b \right) + \log \frac {\mbt}{\mbo} \left[ X_b + \left( 2 \hb - \ghat \right) R_b' \right]\right\};\nonumber \\
\left.  {\cal M}^2_H \right|_{\phi_2 \phi_2} &=&M_Z^2 \sin^2 \beta + m_A^2 \cos^2 \beta
\nonumber \\
&&+ \frac {3 m_t^2} {8 \pi^2} \left\{ \htop \log \frac {\mto \mtt} {m_t^4}- \ghat \log \frac {\mto \mtt} {Q_0^4}
\right. \nonumber \\
&& \left. + g(\mto,\mtt) R_t' \left( \htop R_t' + X_t \right) + \log \frac {\mtt} {\mto} \left[ X_t + \left( 2
\htop - \ghat \right) R_t' \right] \right\} \label{e15b}
\nonumber \\
&+& \frac {3 m_b^2} {8 \pi^2} \left[ g(\mbo,\mbt) R_b \left( \hb R_b - \tanb X_b \right) + \ghat \tanb R_b \log
\frac {\mbt} {\mbo} \right] . \label{e15c}
\end{eqnarray}
where
\be \label{e11} g(m_{1}^2, m_{2}^2) = 2 - \frac {m_{1}^2 + m_{2}^2} {m_{1}^2 - m_{2}^2} \log \frac {m_{1}^2}
{m_{2}^2}, \ee and
 \begin{eqnarray}
 \label{e14}
X_t &=& \frac{ 5 g'^2 - 3 g^2} {12} \cdot \frac {m_{t_L}^2 - m_{t_R}^2} {\mtt - \mto}\,\,;
 X_b = \frac{ g'^2- 3 g^2} {12} \cdot \frac {m_{t_L}^2 - m_{b_R}^2} {\mbt - \mbo}\,,
\label{e14a} \\
R_t &=& \frac {A_{t^\prime}^2 \cotb + A_t  A_{t^\prime}} { \mtt - \mto} \,\,;
 R_t'= \frac { A_t^2 + A_t  A_{t^\prime} \cotb} { \mtt - \mto} \,
\label{e14b} \\
R_b &=& \frac {A_{b^\prime}^2 \tanb + A_b  A_{b^\prime}} { \mbt - \mbo}\,\,;
 R_b'= \frac {A_b^2 + A_b  A_{b^\prime} \tanb} { \mbt - \mbo }\, .
\label{e14c}
 \end{eqnarray}
It is known that, the two-loop corrections to Higgs boson mass are reduced at the renormalization scale $Q_0 = m_t$
hence our choice hereon.
\begin{figure}[htb]
\begin{center}
\vspace{0.0cm}
        \includegraphics[height=4.5 cm,width=7cm]{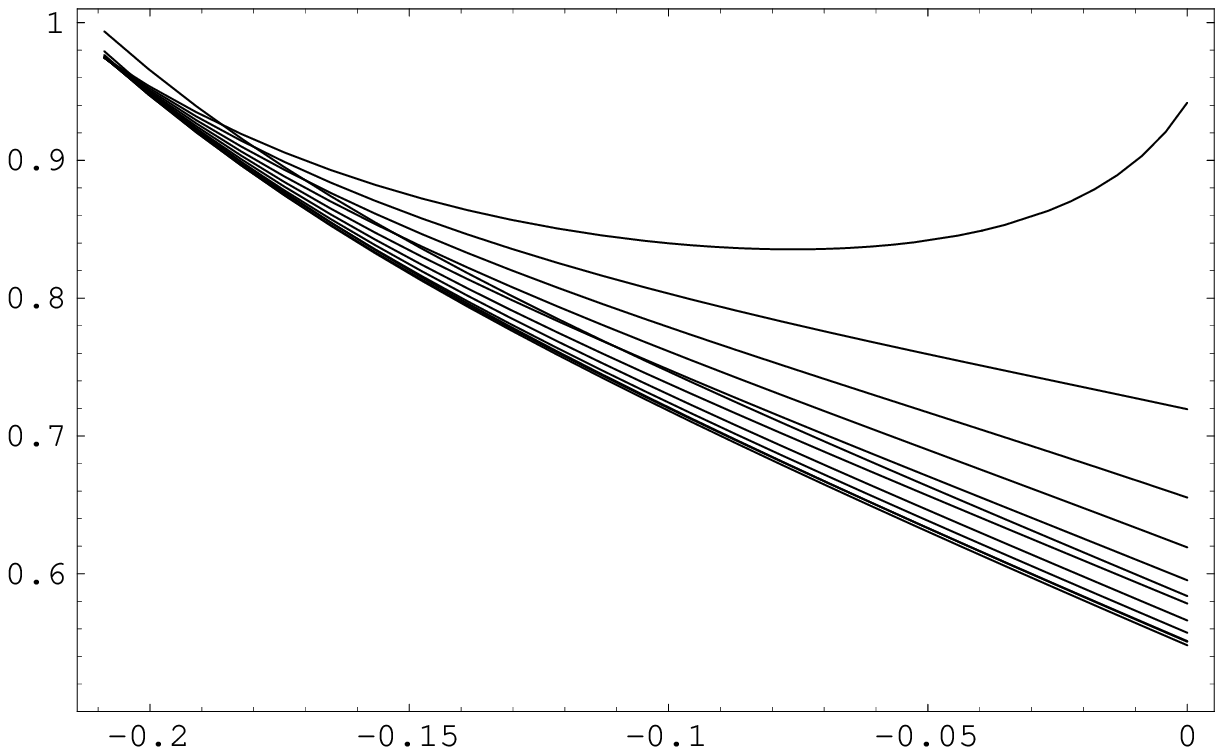}\hspace{0.2cm}
            \includegraphics[height=4.5 cm,width=7cm]{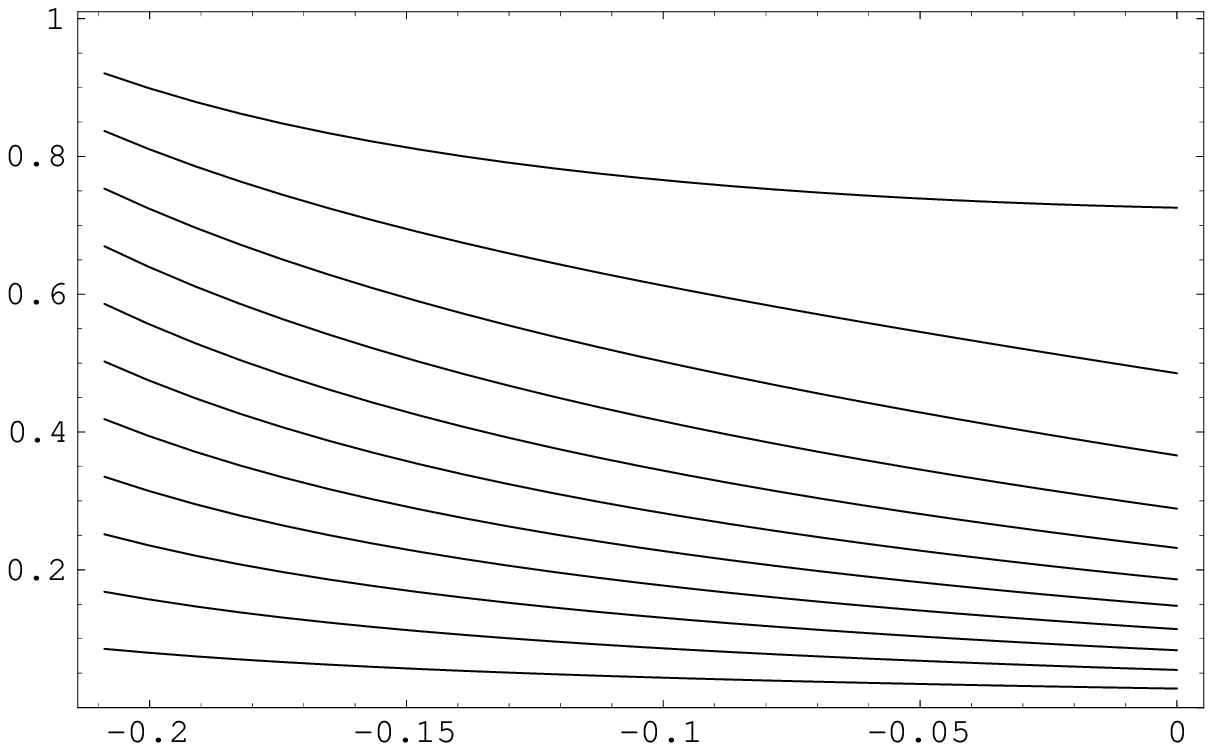}
    \vspace{0cm}
    \caption[]{Scale dependence  of the couplings $h_t$ (left) and of $h_b$ (right)
    for different choices of $tan \beta$ $\epsilon$ [5,55].
    In both of the figures topmost curves correspond to $tan \beta$=55.}
\begin{picture}(0,0)(0,0)
 \put(-120,45){t} 
 \put(-217,115){{$h_t$}} 
  \put(110,45){t} 
 \put(-3,115){{$h_b$}} 
  \label{fighthb}
 \end{picture}
    \end{center}
    \vspace{-1.0cm}
\end{figure}
To give a concrete example of NHSSM benchmark we now list mass predictions of the model for low $tan\beta$ with
the input parameters; $\bar{m}_t(t_Z) = 170,\,\bar{m}_b(t_Z) = 2.92$ and $\bar{m}_\tau(t_Z) = 1.777$ GeV and
take the GUT boundary values of soft terms as the following set
\begin{equation}
M=160,\, m_0=683,\, \mu^\prime_0=400,\, A_0=800,\, A_0^\prime=1000,\,m_{30}=430\,
\end{equation}
which brings the following predictions
\begin{eqnarray}
&&m_{\tilde{t}_1}(t_Z)=291,\,m_{\tilde{t}_2}(t_Z)=626,\,\nonumber\\
&&m_{\tilde{b}_1}(t_Z)=600,\,m_{\tilde{b}_2}(t_Z)=791,\,\nonumber\\
&&m_{\tau_1}(t_Z)=683,\,m_{\tau_2}(t_Z)=695,\,\nonumber\\
&&m_{\chi^0_{1,2,3,4}}(t_Z)=63,\,120,\,392,\,407,\,\\
&&m_{\chi^{\pm}_{1,2}}(t_Z)=119\,,407,\nonumber\\
&&m_{A^0}(t_Z)=289,\,m_{H^\pm}(t_Z)=300,\,\nonumber\\
&&m_{H^0}(t_Z)= 291,\,m_{h^0}(t_Z)_{corrected}=123,\,\nonumber
\end{eqnarray}
where all masses are given in GeV.
\begin{figure}[htb]
\begin{center}
 \vspace{0.0cm}
    \includegraphics[height=10 cm,width=16cm]{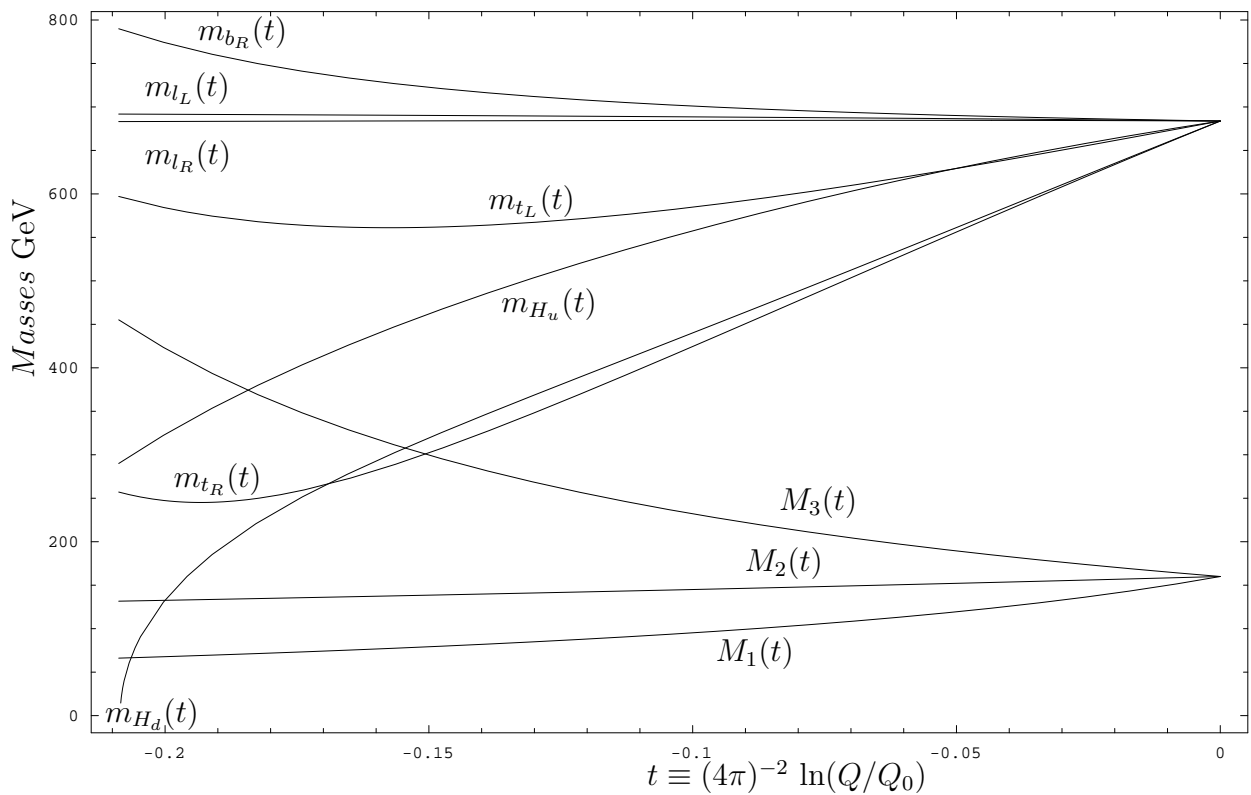}
    \caption[]{A sample plot of some of the soft  terms versus scale in the NHSSM with the input parameters given in the text.}
    \vspace{-.5cm}
 \begin{picture}(0,0)(0,0)
 \put(26,78){$M_1(t)$}
 \put(37,112){$M_2(t)$}
 \put(50,135){$M_3(t)$}
 \put(-205,55){$m_{H_d}(t)$}
  \put(-179,144){$m_{t_R}(t)$}
  \put(-60,249){$m_{t_L}(t)$}
    \put(-55,210){$m_{H_u}(t)$}
     \put(-170,313){$m_{b_R}(t)$}
   \put(-190,266){$m_{l_R}(t)$}
    \put(-190,292){$m_{l_L}(t)$}
 \put(-240,185){\rotate{$Masses $ GeV}}
  \put(0,31){$t\equiv (4
\pi)^{-2}\, \ln({Q}/{Q_0}) $}
    \label{Allmasses}
 \end{picture}
    \end{center}
    \vspace{-.5cm}
\end{figure}

Since NHSSM covers MSSM any prediction of the classical MSSM results can be reproduced in non-holomorphic case
with the appropriate boundaries. But the extension enriches us with more opportunities. What it is important
here is the degree of freedom offered by NHSSM. As it was stressed in \cite{drtjones} for $m_0\gg M$ it turns
out that $|\mu|<0.4~M$ in the MSSM whereas in the NHSSM this constrained is significantly relaxed. Note that in
our example we assumed all soft terms as if they are real and positive without considering any specific model,
whereas one can study i.e $A_0=-M$ which arises in certain string inspired models.  Under the light of these
observations, it should be stated that, NH extension of the MSSM not only covers the classical MSSM but also
offers novel features that can ease the shortcomings of the MSSM, which should be studied in more detail.
Actually, in addition to  LEP limits on the SUSY mass spectrum, one should also deal with the constraints from
$b\rightarrow s \gamma$ decay (as we do in next subsection) and the lower limit on the lifetime of the universe,
which requires the dark matter density from the LSP not to close the universe on itself \cite{kazakov}.

\subsection{$b\rightarrow s \gamma$ Decay}
\label{bsgam} Presently, one of the most accurate observables which can severely constrain the soft masses is
the branching ratio for the rare radiative inclusive $B$ meson decay, $B\rightarrow X_s \gamma$. The main
interest in this decay drives from the genuine perturbative nature of the problem and also from the striking
agreement between the experiment and the SM prediction. Indeed, the measurements of the branching ratio at CLEO,
ALEPH and BELLE gave the combined result \cite{bsgamexp}
\begin{eqnarray}
\label{exp}
\mbox{BR}\left(B\rightarrow X_s \gamma\right)= \left(3.11 \pm 0.42 \pm
0.21 \right)\times 10^{-4}
\end{eqnarray}
whose agreement with the next--to--leading order (NLO) standard model (SM) prediction \cite{misiak}
\begin{eqnarray}
\label{sm} \mbox{BR}\left(B\rightarrow X_s \gamma\right)_{\mbox{SM}} = \left(3.29 \pm 0.33\right)\times
10^{-4}\:
\end{eqnarray}
is manifest though the inclusion of the nonperturbative effects can modify the result
slightly \cite{kagan1}. That the experimental result (\ref{exp}) and the
SM prediction (\ref{sm}) are in good  agreement shows that the ``new
physics" should lie well above the electroweak scale unless certain
cancellations occur.

The branching ratio for  $B\rightarrow X_s \gamma$ has been computed up to NLO precision in
the MSSM \cite{nlosusy}. The $W$ boson and charged Higgs contributions are of the same sign
and thus the chargino--stop loop is expected to moderate the branching ratio so as to respect
the experimental bounds. The recent measurements of $\mbox{BR}\left(B\rightarrow X_s \ell^+\ell^-\right)
$ \cite{gambino} imply that the sign of the total $b\rightarrow s \gamma$ amplitude must be
same as in the SM. This eliminates part of the supersymmetric parameter space in which the total
amplitude approximately equals negative of the SM prediction. In spite of these, however, the present
experimental results do not exclude stop masses around a few $M_Z$ as long as $A_t$ and $A_t^{\prime}$
are of opposite sign \cite{nlosusy}.

\begin{figure}[htb]
\begin{center}
\vspace{0.0cm}
        \includegraphics[height=4.5 cm,width=7cm]{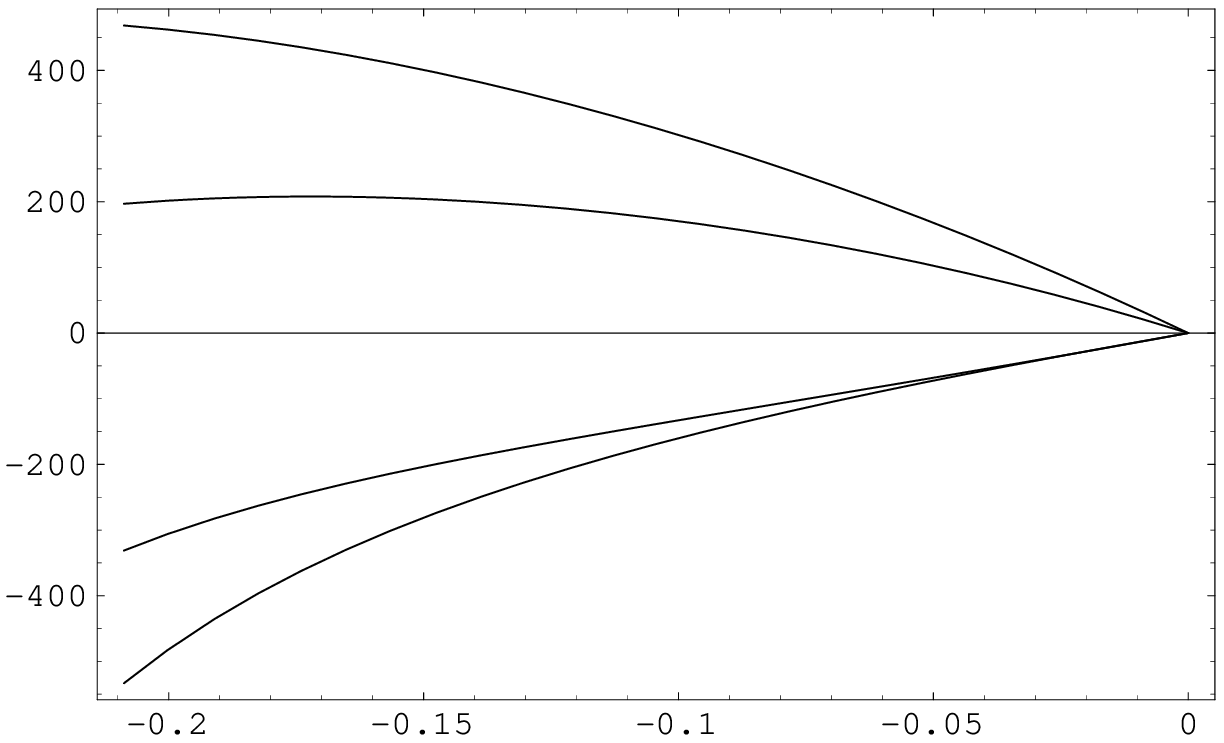}\hspace{0.2cm}
            \includegraphics[height=4.5 cm,width=7cm]{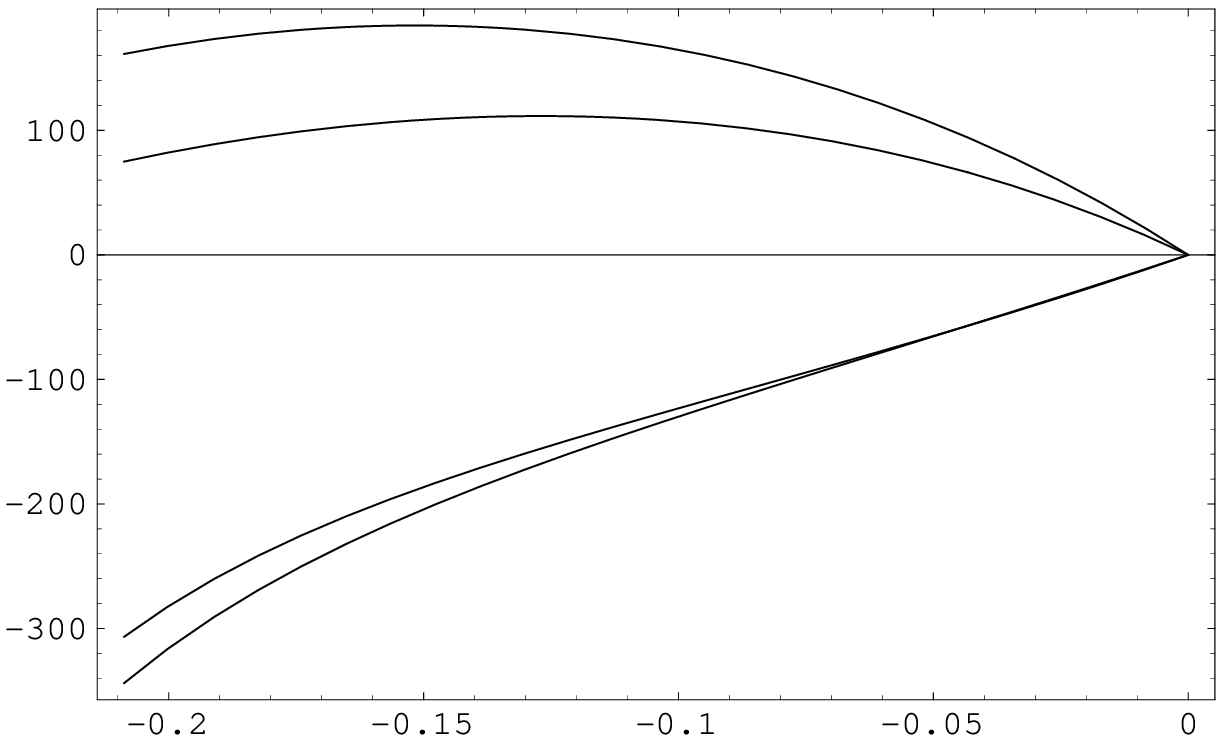}
    \vspace{0cm}
    \caption[]{A sample plot of the scale dependence of the trilinear couplings
     for $tan \beta$=5 (left), $tan \beta$=50  (right) with $A_0=A_0^\prime$=0,
     $M$=150 GeV and $\mu^\prime$=1000 GeV, which show a candidate region where $A_t$ and $A_t^{\prime}$
are of opposite sign.} \vspace{-.5cm}
\begin{picture}(0,0)(0,0)
 \put(-101,45){t} 
 \put(-183,95){{$A_t$}} 
  \put(-161,71){{$A_b$}} 
  \put(-180,165){{$A_{t^\prime}$}} 
   \put(-180,135){{$A_{b^\prime}$}} 
  \put(110,45){t} 
 \put(32,90){{$A_t$}} 
  \put(38,65){{$A_b$}} 
  \put(30,164){{$A_{t^\prime}$}} 
   \put(30,145){{$A_{b^\prime}$}} 
    \label{figAtb}
 \end{picture}
    \end{center}
    \vspace{.5cm}
\end{figure}
To accommodate differing signs of trilinear couplings in the NHSSM we present another example using the
following input parameter
\begin{equation}
M=200,\, m_0=787,\, \mu^\prime_0=400,\, A_0=900,\, A_0^\prime=-1500,\,m_{30}=414\,
\end{equation}
which yields the following predictions
\begin{eqnarray}
&&m_{\tilde{t}_1}(t_Z)=362,\,m_{\tilde{t}_2}(t_Z)=728,\,\nonumber\\
&&m_{\tilde{b}_1}(t_Z)=711,\,m_{\tilde{b}_2}(t_Z)=930,\,\nonumber\\
&&m_{\tau_1}(t_Z)=787,\,m_{\tau_2}(t_Z)=801,\,\nonumber\\
&&m_{\chi^0_{1,2,3,4}}(t_Z)=79,\,150,\,392,\,409\,\\
&&m_{\chi^{\pm}_{1,2}}(t_Z)=149\,,408,\nonumber\\
&&m_{A^0}(t_Z)=299,\,m_{H^\pm}(t_Z)=310,\,\nonumber\\
&&m_{H^0}(t_Z)=301,\,m_{h^0}(t_Z)_{corrected}=120,\,\nonumber
\end{eqnarray}
 here again  all masses are given in GeV.

\section{Renormalization Group Invariants in the MSSM and NHMSSM:
A Comparative Analysis}
Renormalization Group Invariants (RGIs), which can be used to relate measurements at the electroweak scale
to physics at ultra high energies provide important information about high scale physics due to the scale
invariance of the quantities under concern \cite{kobayashi1,durmus}.
Since the coupled nature of the RGEs disturbs analytical solutions
it would be beneficial to know if one can construct certain invariants that  give relations among
the spectrum of supersymmetric particles.
Indeed, RG invariants may provide a direct,
accurate way of testing the internal consistency of the model and
determine the mechanism which breaks the supersymmetry. Such
quantities prove highly useful not only for projecting the experimental data to high energies but also for
deriving certain sum rules which enable fast consistency checks of the model.
Assume there is a measurement which tells a specific relation   between  some of the soft masses, then,
it can be easily probed whether this relation  survives at different scales or not, with the help of
 scale independent relations, which in turn shows the way how SUSY is broken.

In this part we will discuss RG
invariant observables in supersymmetry with non-holomorphic soft terms and
compare with existing MSSM results with the assumption that there is no flavor mixing
and soft terms obey the universality condition mentioned previosly. Nevertheless,  it should be kept in mind that
we study one-loop RGIs  which differs  when R parity or higher loop effects are taken into account.

To begin with, note that lagrangian of the NHSSM (\ref{lagran}) has parameters  defined at a
specific mass scale $Q$ which can physically range from the
electroweak scale $Q=M_Z$ (the IR end) up to some high energy
scale $Q=Q_0$ (the UV end). For determining the scale dependencies
of the parameters the RGEs are to be solved with proposed boundary
conditions either at IR or UV. In what follows we will
write them in terms of the dimensionless variable $t\equiv (4
\pi)^{-2}\, \ln({Q}/{Q_0}) $, and solve for the parameters  in
terms of their UV scale values by taking into account the fact
that the gauge and Yukawa (at a given $\tan\beta$) couplings are
already known at IR end.

We should deal with the rigid parameters in both of the models as a first step.
The RGEs for gauge and
Yukawa couplings form a coupled set of first order differential
equations and can be found elsewhere (i.e. see \cite{rge}). Now one can solve them at any scale at one loop
order without resorting to other model parameters.
 However,
expanding this set of equations by including the RGE of the $\mu^\prime$
parameter one finds that
\begin{eqnarray}
\label{I1} I_1 = \mu^\prime\, \left(\frac{g_2^9\, g_3^{256/3}}{h_t^{27}\,
h_b^{21}\, h_{\tau}^{10}\, g_1^{73/33}}\right)^{1/61}
\end{eqnarray}
is a one-loop RG-invariant.
For the classical MSSM invariant $\mu^\prime\rightarrow\mu$ substitution suffices (MSSM was also mentioned in \cite{kobayashi1}).
Here the powers of the Yukawa and gauge couplings
follow from group-theoretic factors appearing in their RGEs.
This invariant provides an explicit solution for the $\mu^\prime$ parameter
\begin{eqnarray}
\label{mueqn} \mu^\prime(t) = \mu^\prime(0)\,
\left(\frac{h_t(t)}{h_t(0)}\right)^{\frac{27}{61}}\,
\left(\frac{h_b(t)}{h_b(0)}\right)^{\frac{21}{61}}\,
\left(\frac{h_{\tau}(t)}{h_{\tau}(0)}\right)^{\frac{10}{61}}\,
\left(\frac{g_3(0)}{g_3(t)}\right)^{\frac{256}{183}}\,
\left(\frac{g_2(0)}{g_2(t)}\right)^{\frac{9}{61}}\,
\left(\frac{g_1(t)}{g_1(0)}\right)^{\frac{73}{2013}}
\end{eqnarray}
once the scale dependencies of gauge and Yukawa couplings are
known either via direct integration or via approximate solutions
 the RGE of the $\mu^\prime$ parameter
involves only the Yukawa couplings, $g_2$ and $g_1$ though this
explicit solution bears an explicit dependence on $g_3$. This
follows from the RGEs of the Yukawa couplings. One of the most interesting sides
 of this invariant is that weights of all gauge and Yukawa couplings is made obvious.
  With this equation one can determine the amount of fine tuning to satisfy Z mass boundary
  (see reference cite{durmus} for a detailed discussion on this issue).
Another  by-product of the invariant $I_1$ is that the phase of
the $\mu$ parameter is an RG invariant.
Since the contribution of higher order loop effects affect invariance relation of (\ref{I1}) $\sim 2-3 \%$; an effect likely to get embodied in the
experimental errors encourages us to work at one-loop order. On the other hand, once the flavor
mixings in Yukawa matrices are switched on there is no obvious
invariant like (\ref{I1}) even at one loop order.

We continue our analysis with the construction of the RG invariants
of the soft parameters of the theory. Of this sector,
a well-known RG invariant is the ratio of the gaugino masses to fine structure constants
\begin{eqnarray}
\label{I2} I_2  = \frac{M_a}{g_a^2}
\end{eqnarray}
with one-loop accuracy. This very invariant guarantees that
\begin{eqnarray}
\label{maeqn} M_a(t) = M_a(0) \left(\frac{g_a(t)}{g_a(0)}\right)^2
\end{eqnarray}
so that knowing two of the gaugino masses at $Q=M_Z$ suffices to know the third -- an important aspect to check
directly the minimality of the gauge structure using the experimental data. Related with this invariant it is
useful to state the well known mass ratios $M_3(t_Z)/M_2(t_Z)=3.46$ and $M_2(t_Z)/M_1(t_Z)=1.99$ at one-loop
order. The invariant (\ref{I2}) pertains solely to the gauge sector of the theory; it is completely immune to
non-gauge parameters. At two loops $I_2$ is no longer an invariant; it is determined by a linear combination of
gaugino masses and trilinear couplings. Combining (\ref{mueqn}) and (\ref{maeqn}) one concludes that the
chargino and neutralino sectors of the theory are connected to the UV scale via the gauge and Yukawa couplings
alone. The equation (\ref{maeqn}) suggests that $M_3(t_Z)/M_3(0)$ is much larger $M_{1,2}(t_Z)/M_{1,2}(0)$ due
to asymptotic freedom, and these coefficients stand still whatever happens in the sfermion and Higgs sectors of
the theory.

A by-product of the invariant (\ref{I2}) is that the phases of the
gaugino masses are RG invariants (like that of the $\mu$
parameter). However, this is correct only at one-loop level; at
two loops the phases of the trilinear couplings disturb the
relation between IR and UV phases of the gaugino masses.

Another invariant of mass dim-1 is related with the B parameter for which we obtain:
\begin{eqnarray}
\label{I3} I_3 &=& B - \frac{27}{61} A_t - \frac{21}{61} A_b -
\frac{10}{61} A_{\tau} - \frac{256}{183} M_3 - \frac{9}{61} M_2 +
\frac{73}{2013} M_1\nonumber\\
&+&c_1 A_t^\prime +c_2 A_b^\prime+
c_3 A_{\tau}^\prime-(c_1+c_2+c_3)\mu^\prime,
\end{eqnarray}
with arbitrary coefficients $c_i$ such that in the limit
$A_{t,b,\tau}^{\prime},\mu^\prime\rightarrow \mu$ it reproduces the well known MSSM invariant
 which can  be expressed in terms of other parameters
\begin{eqnarray}
\label{beqn}
B(t) &=& B(0)+\frac{27}{61} \left(A_t(t)-A_t(0)\right) + \frac{21}{61} \left(A_b(t)-A_b(0)\right)
+\frac{10}{61} \left(A_{\tau}(t)-A_{\tau}(0)\right)\\
&+&\frac{256}{183} M_3(0) \left(\frac{g_3(t)^2}{g_3(0)^2} - 1\right)
+ \frac{9}{61} M_2(0) \left( \frac{g_2(t)^2}{g_2(0)^2} - 1\right)
-\frac{73}{2013} M_1(0) \left( \frac{g_1(t)^2}{g_1(0)^2} - 1\right).\nonumber
\end{eqnarray}

Concerning mass dimension-2 terms we obtain a general invariant relation in the NHSSM by brute force as follows
\begin{eqnarray}
\label{I4}
I_4&=&\left(\frac{c_1}{6}+\frac{9 c_2}{16}+\frac{c_3}{2}+\frac{c_4}{2}\right) m_{H_u}^2(t)
+\left(\frac{-c_1}{6}+\frac{3 c_2}{16}-\frac{c_3}{2}-\frac{c_4}{2}\right) m_{H_d}^2(t)\,\nonumber\\
&+&\left(\frac{c_1}{2}-\frac{9 c_2}{16}-\frac{c_3}{2}+\frac{3 c_4}{2}\right) m_{t_L}^2(t)
+\left(\frac{-c_1}{2}-\frac{9 c_2}{16}-\frac{c_3}{2}-\frac{3 c_4}{2}\right) m_{t_R}^2(t)\,\nonumber\\
&+&\left(\frac{c_1}{6}-\frac{3 c_2}{16}+\frac{c_3}{2}-\frac{3 c_4}{2}\right) m_{l_L}^2(t)+c_3 m_{b_R}^2(t)+c_4 m_{l_R}^2(t)\,\nonumber\\
&-&\left(\frac{c_1}{33}+\frac{c_2}{44}\right)M_1^2(t)+c_1 M_2^2(t) +c_2 M_3^2(t)\,\nonumber\\
&+&c_5 A_t^{\prime 2}(t)+c_6 A_b^{\prime 2}(t)+c_7 A_\tau^{\prime 2}(t)-\left(\frac{3 c_2}{4}+c_5+c_6+c_7 \right)\mu^{\prime 2}(t).
\end{eqnarray}
where $c_i$ are arbitrary constants.
To visualize our results lets set all coefficient to zero but $c_{5,6,7}$ we then obtain
\begin{equation}
c_5 A_t^{\prime 2}(t)+c_6 A_b^{\prime 2}(t)+c_7 A_\tau^{\prime 2}(t)-(c_5+c_6+c_7)\mu^{\prime 2}(t),
\end{equation}
which is obviously invariant in the limit $A_{t,b,\tau}^{\prime},\mu^\prime\rightarrow \mu$. Note that using this
limiting case one can obtain another invariant,
 when supplemented with $m_{H_{u,d}}^2(t)\rightarrow m_{H_{u,d}}^2(t)+\mu^2(t)$ brings the most general form of MSSM invariant
mass of dim-2. In the cases when we relax these substitutions  we obtain more general structures.
Now we vary the coefficients of various soft masses for constructing invariants in terms of $M_i$ and $\mu$ parameters.
Using this freedom, when we set $ c_1= -3,\,c_4= 1$ and all other coefficients to zero, we get
\begin{equation}
I_5= -2 m_{lL}^2(t)+m_{lR}^2(t)+3 \left|M_2(t)\right|^2-\frac{1}{11}\left|M_1(t)\right|^2
\end{equation}
and similarly various patterns of  the coefficients give rise to
\begin{eqnarray}
I_6&=& m_{H_u}^2(t)-\frac{3}{2} m_{t_R}^2(t)+\frac{4}{3} \left|M_3(t)\right|^2+\frac{3}{2}\left|M_2(t)\right|^2 -\frac{5}{66}\left|M_1(t)\right|^2-\left|\mu^\prime(t)\right|^2,\,\nonumber\\
I_7&=& m_{H_d}^2(t)-\frac{3}{2} m_{b_R}^2(t)-m_{l_L}^2(t)+\frac{4}{3} \left|M_3(t)\right|^2-\frac{1}{33}\left|M_1(t)\right|^2-\left|\mu^\prime(t)\right|^2,\,\nonumber\\
I_8&=& m_{t_R}^2(t)+ m_{b_R}^2(t)-2 m_{t_L}^2(t)-3 \left|M_2(t)\right|^2+\frac{1}{11}\left|M_1(t)\right|^2,\,\\
I_9&=& m_{H_u}^2(t)+ m_{H_d}^2(t)-3 m_{t_L}^2(t)-m_{l_L}^2(t)\,\nonumber\\
&+&\frac{8}{3} \left|M_3(t)\right|^2-3\left|M_2(t)\right|^2 +\frac{1}{33}\left|M_1(t)\right|^2-2\left|\mu^\prime(t)\right|^2,\,\nonumber\\
I_{10}&=& m_{H_d}^2(t)-\frac{3}{2} m_{b_R}^2(t)-\frac{3}{2} m_{l_L}^2(t)+\frac{1}{4}m_{l_R}^2(t)\nonumber\\
&+&\frac{4}{3} \left|M_3(t)\right|^2-\frac{3}{4}\left|M_2(t)\right|^2 +\frac{1}{132}\left|M_1(t)\right|^2-\left|\mu^\prime(t)\right|^2,\,\nonumber
\end{eqnarray}
which should be compared with the results of (see \cite{durmus}). Clearly, one can construct new invariants by combining the ones presented
here or by varying the coefficients expressed as $c_i$.
 Although the results presented here and the results  of \cite{durmus} coincide a term is observed to be missing in
some of the invariant equations. This stems from  the definitions and frameworks $i.e.$ we work within minimal
supergravity (with non-holomorphic soft terms). Here we confirm the results of \cite{kobayashi1,durmus} in
certain limits and we also generate new invariants.

The general form (\ref{I4}) and the invariants that follow could be very useful
for sparticle spectroscopy \cite{spect} in that they
provide scale-invariant correlations among various sparticle
masses.
\begin{figure}[htb]
\begin{center}
 \vspace{0.0cm}
    \includegraphics[height=5 cm,width=8cm]{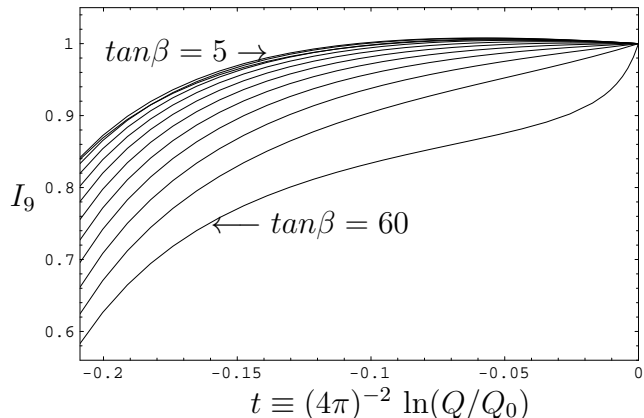}
    \caption[]{Fixed point behavior of the anomal $I_9$ against scale.
    Here we assume same weight for all soft terms ($\sim 40$ GeV) and re-scale the figure (initial value of this invariant is $\sim 32\, TeV^2$).}
 \begin{picture}(0,0)(0,0)
 \put(-50,125){$\longleftarrow tan\beta=60$}
  \put(-90,190){$ tan\beta=5 \rightarrow$}
 \put(-126,135){{$I_9$}}
  \put(-35,57){$t\equiv (4\pi)^{-2}\, \ln({Q}/{Q_0}) $}
   \label{In9fig}
 \end{picture}
    \end{center}
      \vspace{-1cm}
      \end{figure}
      All the invariants presented here show non-anomalous behaviors unless they bear $\mu^\prime$ terms.
       As an example lets take $I_9$ Fig. (\ref{In9fig}), which demonstrates the fixed behavior.
       Notice that while it is scale-dependent, it is still  very useful since its dependency is very soft.
However, notice that they are obtained without noting flavor mixing and in the mSUGRA framework.
 Nevertheless, using them one can ($i$) test the internal consistency of
the model while fitting to the experimental data; ($ii$)
rehabilitate poorly known parameters supplementing the
well-measured ones; ($iii$) determine what kind of supersymmetry
breaking mechanism is realized in Nature; and finally ($iv$)
separately examine the UV scale configurations of the trilinear
couplings as they do not explicitly contribute to the invariants.

 Consequently,
if one single invariant is measured then all are done, and in case
the experimental data prefer a certain correlation pattern among
the invariants then the corresponding UV scale model is preferred.
In this sense, rendering unnecessary the RG running of individual
sparticle masses up to the messenger scale, the invariants speed
up the determination of what kind of supersymmetry breaking
mechanism is realized in Nature.

\section{Conclusion}

It is important to explore the features of MSSM and its extensions as general as possible.
This will be clear as experimental data accumulates about the masses of all predicted particles,
 and for the time being it should be calculated at low energies using the RGEs. For that aim NHSSM
offers novel opportunities which should be studied in more detail. Compared with its enrichments,
 there are not enough papers in the literature about the phenomenological consequences of the NHSSM.
 So we try to cover this issue from many sides.
  Because we do not know the mechanism of
supersymmetry breaking, extensions of the MSSM should be taken seriously to ease the shortcomings of the MSSM.
In this paper we explored the main features of NHSSM with minimal particle content and  observe that, in addition
to mimic the reactions of the MSSM (like gauginos or Yukawa couplings), NHSSM offers interesting opportunities.
 Even, under certain
assumptions, it is possible to completely get rid of famous $\mu$ problem in the NHSSM, and this corresponds to
two special turning points in low and high $tan\beta$ regimes, which is not possible in classical MSSM. The
price that must be paid is, facing additional primed trilinear coupling and fine tuning of parameters for GUT
boundaries.

One of the main results of this work is to present semi-analytic  solutions of RGEs of NHSSM which enables one
to study the phenomenology in detail. Using the solutions presented here one can investigate the reaction of the
NHSSM deeper. Notice that the solutions presented in the Appendices have nonzero phases which should be used to
go deeper in the phenomenology.

 Another result is to present a general form of RGIs which can be used to derive new relations
  in addition to those existing in the literature. We observed that by using existing RGEs one can construct
  RGIs with a simple computer code which indeed offers a very practical way of handling the equations.
  These invariants turn out to be highly useful in making otherwise
indirect relations among the parameters manifest.  Moreover, they serve as efficient tools for performing fast
consistency checks for deriving poorly known parameters from known ones in course of fitting the model to
experimental data, and  for probing the mechanism that breaks the supersymmetry.

\section{Acknowledgement}
We are grateful to D. A. Demir  for invaluable discussions. One of the authors (L. S.) would like to express his
gratitude to the  Izmir Institute of Technology where part of this work has been done.
\newpage
\appendix
\section{Explicit form of RGEs of the NHSSM}
\label{appRGEs}
For the NHSSM  one-loop renormalization group equations can be found in \cite{drtjones}
we also present here for the sake of completeness.
\begin{eqnarray}
\label{RGEs1} \beta_{m_{H_d}^2}&=& 2 h_{\tau}^2 ( m_{H_d}^2+ A_{\tau}^2+ m_{l_L}^2+m_{l_R}^2)+6 h_{b}^2 ( m_{H_d}^2+ A_{b}^2+ m_{t_L}^2+ m_{b_R}^2) \nonumber \\
&+&6 h_{t}^2 \ A_{t^\prime}^2-8 C_H \mu^{\prime 2}-6 g_2^2 M_2^2-2 g^{\prime 2} M_1^2,\\ \nonumber\\
\label{RGEs2}
\beta_{m_{H_u}^2}&=& 6 h_{t}^2 ( m_{H_u}^2+ A_{t}^2+ m_{t_L}^2+ m_{t_R}^2)+2 h_{\tau}^2\ A_{\tau^\prime}^2+6 h_{b}^2 A_{b^\prime}^2  \nonumber \\
&-&8 C_H \mu^{\prime 2}-6 g_2^2 \ M_2^2-2 g^{\prime 2} M_1^2, \\ \nonumber\\
\label{RGEs3} \beta_{m_3^2}&=& ( h_{\tau}^2+3 h_{b}^2+3 h_{t}^2) m_3^2+2 h_{\tau}^2 \ A_{\tau^\prime}
 A_{\tau}+6 h_{b}^2 A_{b^\prime} A_{b}+6 h_{t}^2 \ A_{t^\prime} A_{t} \nonumber\\
&-&4 C_H m_3^2+6 g_2^2 \mu\prime M_2+2 \ g^{\prime 2} M_1 \mu^\prime , \\ \nonumber\\
\label{RGEs4}
\beta_{\mu^\prime}&=& ( h_{\tau}^2+3 h_{b}^2+3 h_{t}^2-4 C_H) \mu^\prime , \\ \nonumber\\
\beta_{A_{\tau^\prime}}&=& ( h_{\tau}^2-3 h_{b}^2+3 h_{t}^2) A_{\tau^\prime}+6 h_{b}^2 A_{b^\prime}+(4 A_{\tau^\prime}-8 \mu^\prime) C_H , \\ \nonumber\\
\label{RGEs5}
\beta_{A_{\tau}}&=& 8 h_{\tau}^2 A_{\tau}+6 h_{b}^2 A_{b}+6 g_2^2 M_2+6 \ g^{\prime 2} M_1 , \\ \nonumber\\
\label{RGEs6}
\beta_{A_{b^\prime}}&=& (- h_{\tau}^2+3 h_{b}^2+ h_{t}^2) A_{b^\prime}+2 A_{\tau^\prime} h_{\tau}^2-2 h_{t}^2 ( A_{t^\prime}-2 \mu^\prime)+(4 A_{b^\prime}-8 \mu^\prime) C_H , \\ \nonumber\\
\beta_{A_{b}}&=& 2 h_{\tau}^2 A_{\tau}+12 h_{b}^2 A_{b}+2 h_{t}^2 A_{t}+\frac{32}{3} g_3^2 M_3+6 g_2^2 M_2+\frac{14}{9} g^{\prime 2} M_1, \\ \nonumber\\
\label{RGEs7}
\beta_{A_{t^\prime}}&=& ( h_{\tau}^2+ h_{b}^2 +3 h_{t}^2) A_{t^\prime}-2 A_{b^\prime} h_{b}^2+4 \mu^\prime h_{b}^2+(4 A_{t^\prime}-8 \mu^\prime) C_H, \\ \nonumber\\
\label{RGEs8} \beta_{A_{t}}&=& 2 h_{b}^2 A_{b}+12 h_{t}^2 A_{t}+\frac{32}{3} g_3^2 M_3+6 \ g_2^2
 M_2+\frac{26}{9} g^{\prime 2} M_1 , \\ \nonumber\\
\label{RGEs9} \beta_{m_{t_L}^2}&=& 2 h_{b}^2 ( m_{t_L}^2+ m_{b_R}^2+ m_{H_d}^2+ A_{b^\prime}^2+ A_{b}^2-2 \mu^{\prime 2}) \nonumber \\
&+&2  h_{t}^2 \ ( m_{t_L}^2+ m_{t_R}^2+ m_{H_u}^2+ A_{t^\prime}^2+ A_{t}^2-2 \mu^{\prime 2}) \nonumber\\
&-&\frac{32}{3} g_3^2 M_3^2-6 g_2^2 M_2^2-\frac{2}{9} g^{\prime 2} M_1^2 , \\ \nonumber\\
\label{RGEs10} \beta_{m_{t_R}^2}&=& 4 h_{t}^2 ( m_{t_L}^2+ m_{t_R}^2+ m_{H_u}^2+ A_{t^\prime}^2+ A_{t}^2-2 \mu^{\prime 2})-\frac{32}{3} g_3^2 M_3^2-\frac{ 32 }{ 9 } \ g^{\prime 2} M_1^2 , \\ \nonumber\\
\label{RGEs11} \beta_{m_{b_R}^2}&=& 4 h_{b}^2 ( m_{t_L}^2+ m_{b_R}^2+ m_{H_d}^2+ A_{b^\prime}^2+ A_{b}^2-2 \mu^{\prime 2})-\frac{ 32 }{ 3 } g_3^2 M_3^2-\frac{8}{9} g^{\prime 2} \ M_1^2, \\\ \nonumber\\
\label{RGEs12} \beta_{m_{l_L}^2}&=& 2 h_{\tau}^2 ( m_{l_L}^2+ m_{l_R}^2+ m_{H_d}^2+ A_{\tau^\prime}^2+ A_{\tau}^2-2 \mu^{\prime 2})-6 g_2^2 M_2^2-2 g^{\prime 2} M_1^2, \\ \nonumber\\
\label{RGEs13} \beta_{m_{l_R}^2}&=& 4 h_{\tau}^2 ( m_{l_L}^2+ m_{l_R}^2+ m_{H_d}^2+ A_{\tau^\prime}^2+ A_{\tau}^2-2 \mu^{\prime 2})-8 g^{\prime 2} M_1^2 , \\ \nonumber\\
\label{RGEs14} \beta_{M_i}&=&2  b_i M_i g_i^2 , \end{eqnarray}
here $b_{1,2,3} = (\frac{33}{5}, 1, -3)$, ${g'}^2 = \frac{3}{5} g_1^2$, $C_H = \frac{3}{4}g_2^2 +
\frac{3}{20}g_1^2$, $M_{GUT}=1.4\times 10^{16}$ {\rm {GeV}} and $M_Z\leq Q \leq M_{GUT}$.
By assuming that the  SUSY is broken with non-standard soft terms; we obtained semi-analytic solutions
for all soft terms through the one-loop  RGEs given above and express our results at the electro-weak scale in
terms of GUT scale parameters. Our results are presented for  moderate ($tan \beta$=5) and large ($tan
\beta$=50) choices.
\section{Solutions of mass squared $\&$ trilinear terms in the NHSSM}
\label{appnhssm} Using low ($tan\beta=5$) and high ($tan\beta=50$) values of $tan\beta$, the most general form
of the mass-squared and trilinear terms can be written in terms of boundary conditions of  gauge coupling
unification scale which is roughly $M_{GUT}\sim 10^{17}$ GeV. Notice that our phase convention is to assign
$1,2,3$ and $4$ for $M_{1},M_{2},M_{3}$ and $\mu^\prime$; for other quantities it is obvious and can be inferred
from the multipliers.
\subsection{Low $tan\beta$ regime}
\begin{eqnarray}
m_{H_u}^2(t_Z)&=& 0.000216  A_{b_0}^2 - 1.59 \times 10^{-7}    A_{b_0} A_{\tau_0} cos \phi_{b \tau}
- 0.0000203  A_{b_0} M_{10} cos \phi_{b 1}\nonumber\\
&-&  0.000191  A_{b_0} M_{20} cos \phi_{b 2}
- 0.000857  A_{b_0} M_{30} cos \phi_{b 3}
- 0.00124  A_{b^\prime_0}^2\nonumber\\
&+&  1.73 \times 10^{-6}    A_{b^\prime_0} A_{\tau^\prime_0} cos \phi_{b^\prime \tau^\prime}
  -  0.000563  A_{b^\prime_0}\mu_0^{\prime} cos \phi_{b^\prime 4}
  - 0.0869  A_{t_0}^2\nonumber\\
  &+& 0.0000648  A_{t_0} A_{b_0} cos \phi_{t b}
  -  2.05 \times 10^{-8}    A_{t_0} A_{\tau_0} cos \phi_{t \tau}
  + 0.0109  A_{t_0} M_{10} cos \phi_{t 1}\nonumber\\
  &+&  0.0672  A_{t_0} M_{20} cos \phi_{t 2}
  + 0.302  A_{t_0} M_{30} cos \phi_{t 3}
  - 7.96 \times 10^{-8}    A_{\tau_0}^2 \nonumber\\
  &+&  2.25 \times 10^{-8}    A_{\tau_0} M_{10} cos \phi_{\tau 1}
  + 1.19 \times 10^{-7}    A_{\tau_0} M_{20} cos \phi_{\tau 2}
  +  4.14 \times 10^{-7}    A_{\tau_0} M_{30} cos \phi_{\tau 3}\nonumber\\
  &-& 0.000287  A_{\tau^\prime_0}^2
  -  0.000248  A_{\tau^\prime_0}\mu_0^{\prime} cos \phi_{\tau^\prime,4}
  + 0.105  A_{t^\prime_0}^2 \nonumber\\
  &-& 0.000284  A_{t^\prime_0} A_{b^\prime_0} cos \phi_{t^\prime b^\prime}
  +  2.25 \times 10^{-7}    A_{t^\prime_0} A_{\tau^\prime_0} cos \phi_{t^\prime \tau^\prime}
  +  0.0674  A_{t^\prime_0}\mu_0^{\prime} cos \phi_{t^\prime 4}\nonumber\\
  &+& 0.00106  M_{10}^2
  - 0.0058  M_{10} M_{20} cos \phi_{1 2}
  -  0.0291  M_{10} M_{30} cos \phi_{1 3}\nonumber\\
  &+& 0.187  M_{20}^2
  - 0.206  M_{20} M_{30} cos \phi_{2 3}
  - 2.79  M_{30}^2\nonumber\\
  &+&  0.000217  m_{{b_R}0}^2
  + 0.000217  m_{{H_d}0}^2
  + 0.612  m_{{H_u}0}^2\nonumber\\
  &-& 7.98 \times 10^{-8}    m_{{l_L}0}^2
  - 8. \times 10^{-8}    m_{{l_R}0}^2
  -  0.388  m_{{t_L}0}^2 \nonumber\\
  &-& 0.388  m_{{t_R}0}^2
  + 0.136 \mu_0^{\prime 2},
\,\end{eqnarray}
\begin{eqnarray}
m_{H_d}^2(t_Z)&=& -0.0032  A_{b_0}^2 + 5 \times 10^{-6}  A_{b_0} A_{\tau_0} cos \phi_{b \tau}
+0.00018A_{b_0} M_{10} cos \phi_{b 1} \nonumber\\
&+& 0.0022  A_{b_0} M_{20} cos \phi_{b 2}
+ 0.01  A_{b_0} M_{30} cos \phi_{b 3}
+ 2.9 \times 10^{-6}A_{b^\prime_0}^2\nonumber\\
&-&  2.4 \times 10^{-9 } A_{b^\prime_0} A_{\tau^\prime_0} cos \phi_{b^\prime \tau^\prime}
-  0.00017 A_{b^\prime_0}\mu_0^{\prime} cos \phi_{b^\prime 4}
+0.00008  A_{t_0}^2 \nonumber\\
&+& 0.00058  A_{t_0} A_{b_0} cos \phi_{t b}
- 4.7 \times 10^{-7}    A_{t_0} A_{\tau_0} cos\phi_{t \tau}
- 0.000028  A_{t_0} M_{10} cos \phi_{t 1}\nonumber\\
&-& 0.00029  A_{t_0} M_{20} cos \phi_{t 2}
- 0.0013  A_{t_0} M_{30} cos \phi_{t 3}
- 0.00078  A_{\tau_0}^2\nonumber\\
&+&  0.00018  A_{\tau_0} M_{10} cos \phi_{\tau 1}
+ 0.0005  A_{\tau_0} M_{20} cos \phi_{\tau 2}
-7.9 \times 10^{-6}    A_{\tau_0} M_{30} cos \phi_{\tau 3}\nonumber\\
&+& 5.2 \times 10^{-7} A_{\tau^\prime_0}^2
+ 3.5 \times 10^{-7} A_{\tau^\prime_0}\mu_0^{\prime}cos\phi_{\tau^\prime 4}
- 0.37  A_{t^\prime_0}^2\nonumber\\
&-& 0.00026 A_{t^\prime_0} A_{b^\prime_0} cos \phi_{t^\prime b^\prime}
+ 7.5 \times 10^{-8}    A_{t^\prime_0} A_{\tau^\prime_0} cos \phi_{t^\prime \tau^\prime}
-0.31 A_{t^\prime_0}\mu_0^{\prime} cos \phi_{t^\prime 4}\nonumber\\
&+& 0.037  M_{10}^2
- 0.00013  M_{10} M_{20} cos \phi_{1 2}
- 0.0003  M_{10} M_{30} cos \phi_{1 3}\nonumber\\
&+& 0.48 M_{20}^2
- 0.004  M_{20} M_{30} cos \phi_{2 3}
- 0.026 M_{30}^2\nonumber\\
&-& 0.0032  m_{{b_R}0}^2 +  m_{{H_d}0}^2
+ 0.00029 m_{{H_u}0}^2\nonumber\\
&-& 0.00079  m_{{l_L}0}^2
- 0.00079 m_{{l_R}0}^2
- 0.0029  m_{{t_L}0}^2\nonumber\\
&+& 0.00029 m_{{t_R}0}^2+0.6\mu_0^{\prime 2}, \,\end{eqnarray}
\begin{eqnarray}
m_{t_L}^2(t_Z)&=& -0.00099  A_{b_0}^2
+ 9.8 \times 10^{-7}    A_{b_0} A_{\tau_0} cos \phi_{b \tau}
+ 0.000053  A_{b_0} M_{10} cos \phi_{b 1} \nonumber\\
&+&  0.00068  A_{b_0} M_{20} cos \phi_{b 2}
  + 0.003  A_{b_0} M_{30} cos \phi_{b 3}
  - 0.00041  A_{b^\prime_0}^2  \nonumber\\
  &+& 3.1 \times 10^{-7}    A_{b^\prime_0} A_{\tau^\prime_0} cos \phi_{b^\prime \tau^\prime}
  -  0.00024  A_{b^\prime_0} \mu_0^{\prime} cos \phi_{b^\prime 4}
  - 0.029  A_{t_0}^2 \nonumber\\
  &+& 0.00022  A_{t_0} A_{b_0} cos \phi_{t b}
  -  1.2 \times 10^{-7}    A_{t_0} A_{\tau_0} cos \phi_{t \tau}
  + 0.0036  A_{t_0} M_{10} cos \phi_{t 1} \nonumber\\
  &+&  0.022  A_{t_0} M_{20} cos \phi_{t 2}
  + 0.1  A_{t_0} M_{30} cos \phi_{t 3}
  + 4.9 \times 10^{-7}    A_{\tau_0}^2 \nonumber\\
  &-&  1.3 \times 10^{-7}    A_{\tau_0} M_{10} cos \phi_{\tau 1}
  - 6.4 \times 10^{-7}    A_{\tau_0} M_{20} cos \phi_{\tau 2}
  -  1.9 \times 10^{-6}    A_{\tau_0} M_{30} cos \phi_{\tau 3} \nonumber\\
  &+& 0.000039  A_{\tau^\prime_0}^2
  +  0.000024  A_{\tau^\prime_0}\mu_0^{\prime} cos \phi_{\tau^\prime 4}
  - 0.089  A_{t^\prime_0}^2 \nonumber\\
  &-&  0.00018  A_{t^\prime_0} A_{b^\prime_0} cos \phi_{t^\prime b^\prime}
  +  7.7 \times 10^{-8}    A_{t^\prime_0} A_{\tau^\prime_0} cos \phi_{t^\prime \tau^\prime}
  -  0.08  A_{t^\prime_0}\mu_0^{\prime} cos \phi_{t^\prime 4} \nonumber\\
  &-& 0.0081  M_{10}^2
  - 0.002  M_{10} M_{20} cos \phi_{1 2}
  -  0.0098  M_{10} M_{30} cos \phi_{1 3} \nonumber\\
&+& 0.38  M_{20}^2
  - 0.07  M_{20} M_{30} cos \phi_{2 3}
  + 5.4  M_{30}^2 \nonumber\\
  &-&  0.00099  m_{{b_R}0}^2
  - 0.00099  m_{{H_d}0}^2
  - 0.13  m_{{H_u}0}^2 \nonumber\\
  &+& 4.9 \times 10^{-7}    m_{{l_L}0}^2
  + 4.9 \times 10^{-7}    m_{{l_R}0}^2
  +  0.87  m_{{t_L}0}^2 \nonumber\\
  &-& 0.13  m_{{t_R}0}^2
  + 0.3 \mu_0^{\prime 2},
  \,\end{eqnarray}
\begin{eqnarray}
m_{t_R}^2(t_Z)&=& 0.00014  A_{b_0}^2
- 1.1 \times 10^{-7}    A_{b_0} A_{\tau_0} cos \phi_{b \tau}
- 0.000014  A_{b_0} M_{10} cos \phi_{b 1} \nonumber\\
&-&  0.00013  A_{b_0} M_{20} cos \phi_{b 2}
  - 0.00057  A_{b_0} M_{30} cos \phi_{b 3}
  + 0.00037  A_{b^\prime_0}^2 \nonumber\\
  &-&  3.2 \times 10^{-7}    A_{b^\prime_0} A_{\tau^\prime_0} cos \phi_{b^\prime \tau^\prime}
  +  0.000042  A_{b^\prime_0}\mu_0^{\prime} cos \phi_{b^\prime 4}
  - 0.058  A_{t_0}^2 \nonumber\\
  &+& 0.000043  A_{t_0} A_{b_0} cos \phi_{t b}
  -  1.4 \times 10^{-8}    A_{t_0} A_{\tau_0} cos \phi_{t \tau}
  + 0.0072  A_{t_0} M_{10} cos \phi_{t 1}  \nonumber\\
  &+&  0.045  A_{t_0} M_{20} cos \phi_{t 2}
  + 0.2  A_{t_0} M_{30} cos \phi_{t 3}
  - 5.3 \times 10^{-8}    A_{\tau_0}^2 \nonumber\\
  &+&  1.5 \times 10^{-8}    A_{\tau_0} M_{10} cos \phi_{\tau 1}
  + 8. \times 10^{-8}   A_{\tau_0} M_{20} cos \phi_{\tau 2}
  +  2.8 \times 10^{-6}    A_{\tau_0} M_{30} cos \phi_{\tau 3} \nonumber\\
  &+& 0.000077  A_{\tau^\prime_0}^2
  +  0.000048  A_{\tau^\prime_0}\mu_0^{\prime} cos \phi_{\tau^\prime 4}
  - 0.18  A_{t^\prime_0}^2  \nonumber\\
  &-&  0.000073  A_{t^\prime_0} A_{b^\prime_0} cos \phi_{t^\prime b^\prime}
  +  7.7 \times 10^{-9 }   A_{t^\prime_0} A_{\tau^\prime_0} cos \phi_{t^\prime \tau^\prime}
  -  0.16  A_{t^\prime_0}\mu_0^{\prime} cos \phi_{t^\prime 4} \nonumber\\
  &+& 0.043  M_{10}^2
  - 0.0039  M_{10} M_{20} cos \phi_{1 2}
  -  0.019  M_{10} M_{30} cos \phi_{1 3} \nonumber\\
  &-& 0.2  M_{20}^2
  - 0.14  M_{20} M_{30} cos \phi_{2 3}
  + 4.4  M_{30}^2  \nonumber\\
  &+&  0.00014  m_{{b_R}0}^2
  + 0.00014  m_{{H_d}0}^2
  - 0.26  m_{{H_u}0}^2 \nonumber\\
  &-& 5.3 \times 10^{-8}    m_{{l_L}0}^2
  - 5.3 \times 10^{-8}    m_{{l_R}0}^2
  -  0.26  m_{{t_L}0}^2 \nonumber\\
  &+& 0.74  m_{{t_R}0}^2
  + 0.6 \mu_0^{\prime 2},
\,\end{eqnarray}
\begin{eqnarray}
m_{b_R}^2(t_Z)&=& -0.0021  A_{b_0}^2 + 2.1 \times 10^{-6}    A_{b_0} A_{\tau_0} cos \phi_{b \tau}
+ 0.00012  A_{b_0} M_{10} cos \phi_{b 1} \nonumber\\
&+&  0.0015  A_{b_0} M_{20} cos \phi_{b 2}
  + 0.0066  A_{b_0} M_{30} cos \phi_{b 3}
  - 0.0012  A_{b^\prime_0}^2 \nonumber\\
  &+&  9.5 \times 10^{-7}    A_{b^\prime_0} A_{\tau^\prime_0} cos \phi_{b^\prime \tau^\prime}
  -  0.00053  A_{b^\prime_0}\mu_0^{\prime} cos \phi_{b^\prime 4}
  + 0.000053  A_{t_0}^2 \nonumber\\
  &+& 0.00039  A_{t_0} A_{b_0} cos \phi_{t b}
  -  2.2 \times 10^{-7}    A_{t_0} A_{\tau_0} cos \phi_{t \tau}
  - 0.000019  A_{t_0} M_{10} cos \phi_{t 1} \nonumber\\
  &-&  0.00019  A_{t_0} M_{20} cos \phi_{t 2}
  - 0.00086  A_{t_0} M_{30} cos \phi_{t 3}
  + 1 \times 10^{-6}   A_{\tau_0}^2  \nonumber\\
  &-&  2.7 \times 10^{-7}    A_{\tau_0} M_{10} cos \phi_{\tau 1}
  - 1.4 \times 10^{-6}    A_{\tau_0} M_{20} cos \phi_{\tau 2}\
  -  4.1 \times 10^{-6}    A_{\tau_0} M_{30} cos \phi_{\tau 3} \nonumber\\
  &-& 4.5 \times 10^{-8}    A_{\tau^\prime_0}^2
  +  2.3 \times 10^{-7}    A_{\tau^\prime_0}\mu_0^{\prime} cos \phi_{\tau^\prime 4}
  + 0.00068  A_{t^\prime_0}^2 \nonumber\\
  &-&  0.00029  A_{t^\prime_0} A_{b^\prime_0} cos \phi_{t^\prime b^\prime}
  +  1.5 \times 10^{-7}    A_{t^\prime_0} A_{\tau^\prime_0} cos \phi_{t^\prime \tau^\prime}
  +  0.00038  A_{t^\prime_0}\mu_0^{\prime} cos \phi_{t^\prime 4} \nonumber\\
  &+& 0.017  M_{10}^2
  - 0.000042  M_{10} M_{20} cos \phi_{1 2}
  -  0.0002  M_{10} M_{30} cos \phi_{1 3} \nonumber\\
  &-& 0.0017  M_{20}^2
  - 0.0027  M_{20} M_{30} cos \phi_{2 3}
  + 6.3  M_{30}^2  \nonumber\\
  &+&  1  m_{{b_R}0}^2
  - 0.0021  m_{{H_d}0}^2
  + 0.0002  m_{{H_u}0}^2 \nonumber\\
  &+& 1 \times 10^{-6}    m_{{l_L}0}^2
  + 1 \times 10^{-6}    m_{{l_R}0}^2
  -  0.0019  m_{{t_L}0}^2 \nonumber\\
  &+& 0.0002  m_{{t_R}0}^2
  + 0.0029 \mu_0^{\prime 2},
\,\end{eqnarray}
\begin{eqnarray}
m_{l_L}^2(t_Z)&=& 9.6 \times 10^{-7}    A_{b_0}^2 + 1.9 \times 10^{-6}    A_{b_0} A_{\tau_0} cos\phi_{b \tau}
- 3.1 \times 10^{-7} A_{b_0} M_{10} cos \phi_{b 1} \nonumber\\
&-& 1.3 \times 10^{-6}    A_{b_0} M_{20} cos \phi_{b 2}
  - 1.8 \times 10^{-6}   A_{b_0} M_{30} cos \phi_{b 3}
  - 1.3 \times 10^{-9 }  A_{b^\prime_0}^2 \nonumber\\
  &+&  7.9 \times 10^{-7}    A_{b^\prime_0} A_{\tau^\prime_0} cos \phi_{b^\prime \tau^\prime}
   +  5.4 \times 10^{-7}    A_{b^\prime_0}\mu_0^{\prime} cos \phi_{b^\prime 4}
  - 2.6 \times 10^{-8}    A_{t_0}^2 \nonumber\\
  &-&  1.3 \times 10^{-7}    A_{t_0} A_{b_0} cos \phi_{t b}
  - 1.3 \times 10^{-7}    A_{t_0} A_{\tau_0} cos \phi_{t \tau}
  +  2.5 \times 10^{-8}   A_{t_0} M_{10} cos \phi_{t 1} \nonumber\\ &+&  1.1 \times 10^{-7}    A_{t_0} M_{20} cos \phi_{t 2}
  +  2.1 \times 10^{-7}    A_{t_0} M_{30} cos \phi_{t 3}
  - 0.00079  A_{\tau_0}^2 \nonumber\\
  &+& 0.00018  A_{\tau_0} M_{10} cos \phi_{\tau 1}
  +  0.0005  A_{\tau_0} M_{20} cos \phi_{\tau 2}
  - 1.8 \times 10^{-6}   A_{\tau_0} M_{30} cos \phi_{\tau 3} \nonumber\\
  &-& 0.0004  A_{\tau^\prime_0}^2
  -  0.00032  A_{\tau^\prime_0}\mu_0^{\prime} cos \phi_{\tau^\prime 4}
  + 0.00018  A_{t^\prime_0}^2 \nonumber\\
  &+&  6.9 \times 10^{-8}    A_{t^\prime_0} A_{b^\prime_0} cos \phi_{t^\prime b^\prime}
  +  6.8 \times 10^{-8}    A_{t^\prime_0} A_{\tau^\prime_0} cos \phi_{t^\prime \tau^\prime}
  +  0.0001  A_{t^\prime_0}\mu_0^{\prime} cos \phi_{t^\prime 4} \nonumber\\
  &+& 0.038  M_{10}^2
  - 0.000066  M_{10} M_{20} cos \phi_{1 2}
  +  3.2 \times 10^{-7}    M_{10} M_{30} cos \phi_{1 3} \nonumber\\
  &+& 0.48  M_{20}^2
  + 1.5 \times 10^{-6}    M_{20} M_{30} cos \phi_{2 3}
  + 4. \times 10^{-6}    M_{30}^2 \nonumber\\ &+&   9.7 \times 10^{-7}   m_{{b_R}0}^2
  - 0.00079  m_{{H_d}0}^2
  - 6.6 \times 10^{-8}   m_{{H_u}0}^2 \nonumber\\ &+&  1  m_{{l_L}0}^2
  - 0.00079  m_{{l_R}0}^2
  +  9. \times 10^{-7}    m_{{t_L}0}^2 \nonumber\\ &-& 6.6 \times 10^{-8}    m_{{t_R}0}^2
  + 0.0012 \mu_0^{\prime 2},
\,\end{eqnarray}
\begin{eqnarray}
m_{l_R}^2(t_Z)&=& 1.9 \times 10^{-6}    A_{b_0}^2 + 3.9 \times 10^{-6}    A_{b_0} A_{\tau_0} cos
\phi_{b \tau} - 6.3 \times 10^{-7} A_{b_0} M_{10} cos \phi_{b 1} \nonumber\\ &-& 2.6 \times 10^{-6}    A_{b_0}
M_{20} cos \phi_{b 2}
  - 3.7 \times 10^{-6}    A_{b_0} M_{30} cos \phi_{b 3}
  - 2.6 \times 10^{-9 }   A_{b^\prime_0}^2 \nonumber\\ &+&   1.6 \times 10^{-6}    A_{b^\prime_0} A_{\tau^\prime_0} cos \phi_{b^\prime \tau^\prime}
  +  1.1 \times 10^{-6}    A_{b^\prime_0}\mu_0^{\prime} cos \phi_{b^\prime 4}
  - 5.3 \times 10^{-8}    A_{t_0}^2 \nonumber\\ &-& 2.6 \times 10^{-7}    A_{t_0} A_{b_0} cos \phi_{t b}
  - 2.7 \times 10^{-7}    A_{t_0} A_{\tau_0} cos \phi_{t \tau}
  +  5.1 \times 10^{-8}    A_{t_0} M_{10} cos \phi_{t 1} \nonumber\\ &+&  2.3 \times 10^{-7}    A_{t_0} M_{20} cos \phi_{t 2}
  +  4.1 \times 10^{-7}    A_{t_0} M_{30} cos \phi_{t 3}
  - 0.0016  A_{\tau_0}^2 \nonumber\\ &+&  0.00035  A_{\tau_0} M_{10} cos \phi_{\tau 1}
  +  0.001  A_{\tau_0} M_{20} cos \phi_{\tau 2}
  - 3.7 \times 10^{-6}    A_{\tau_0} M_{30} cos \phi_{\tau 3} \nonumber\\ &-& 0.00081  A_{\tau^\prime_0}^2
  -  0.00064  A_{\tau^\prime_0}\mu_0^{\prime} cos \phi_{\tau^\prime 4}
  + 0.00035  A_{t^\prime_0}^2 \nonumber\\ &+&   1.4 \times 10^{-7}    A_{t^\prime_0} A_{b^\prime_0} cos \phi_{t^\prime b^\prime}
  +  1.4 \times 10^{-7}    A_{t^\prime_0} A_{\tau^\prime_0} cos \phi_{t^\prime \tau^\prime}
  +  0.0002  A_{t^\prime_0}\mu_0^{\prime} cos \phi_{t^\prime 4} \nonumber\\ &+&  0.15  M_{10}^2
    - 0.00013  M_{10} M_{20} cos \phi_{1 2}
  +  6.4 \times 10^{-7}    M_{10} M_{30} cos \phi_{1 3} \nonumber\\ &-& 0.0011  M_{20}^2
  + 2.9 \times 10^{-6}    M_{20} M_{30} cos \phi_{2 3}
  - 0.00019  M_{30}^2 \nonumber\\ &+&   1.9 \times 10^{-6}  m_{{b_R}0}^2
  - 0.0016  m_{{H_d}0}^2
  - 1.3 \times 10^{-7}   m_{{H_u}0}^2 \nonumber\\ &-& 0.0016  m_{{l_L}0}^2
  +  m_{{l_R}0}^2
  +  1.8 \times 10^{-6}    m_{{t_L}0}^2 \nonumber\\ &-& 1.3 \times 10^{-7}    m_{{t_R}0}^2
  + 0.0025 \mu_0^{\prime 2}
\,\end{eqnarray}
\begin{eqnarray}
m_{3}^2(t_Z)&=& 0.00012  A_{b^\prime_0} A_{t_0} cos \phi_{b^\prime t} + 1.7 \times 10^{-6}
A_{b^\prime_0} A_{\tau_0} cos \phi_{b^\prime \tau} +  0.000069  A_{b^\prime_0} M_{10} cos \phi_{b^\prime 1}
\nonumber\\ &+&  0.0008  A_{b^\prime_0} M_{20} cos \phi_{b^\prime 2}
  +  0.0036  A_{b^\prime_0} M_{30} cos \phi_{b^\prime 3}
  + 1.5 \times 10^{-6}    A_{\tau^\prime_0} A_{b_0} cos \phi_{\tau^\prime b} \nonumber\\ &-& 1.2 \times 10^{-7}    A_{\tau^\prime_0} A_{t_0} cos \phi_{\tau^\prime t}
  -  0.00052  A_{\tau^\prime_0} A_{\tau_0} cos \phi_{\tau^\prime \tau}
  +  0.000054  A_{\tau^\prime_0} M_{10} cos \phi_{\tau^\prime 1} \nonumber\\ &+&  0.00015  A_{\tau^\prime_0} M_{20} cos \phi_{\tau^\prime 2}
  -  2.4 \times 10^{-6}    A_{\tau^\prime_0} M_{30} cos \phi_{\tau^\prime 3}
  - 0.00017  A_{t^\prime_0} A_{b_0} cos \phi_{t^\prime b} \nonumber\\ &-& 0.27  A_{t^\prime_0} A_{t_0} cos \phi_{t^\prime t}
  + 1.9 \times 10^{-7}    A_{t^\prime_0} A_{\tau_0} cos \phi_{t^\prime \tau}
  +  0.015  A_{t^\prime_0} M_{10} cos \phi_{t^\prime 1} \nonumber\\ &+&  0.092  A_{t^\prime_0} M_{20} cos \phi_{t^\prime 2}
  +  0.39  A_{t^\prime_0} M_{30} cos \phi_{t^\prime 3}
  - 0.051  M_{10}^2 \nonumber\\ &-& 0.51  M_{20}^2
  + 0.96  m_{30}^2
  -  0.00044 \mu_0^{\prime} A_{b_0} cos \phi_{4 b} \nonumber\\ &-& 0.098 \mu_0^{\prime} A_{t_0} cos \phi_{4 t}
  -  0.00024 \mu_0^{\prime} A_{\tau_0} cos \phi_{4 \tau}
  + 0.0079 \mu_0^{\prime} M_{10} cos \phi_{4 1} \nonumber\\ &+&   0.052 \mu_0^{\prime} M_{20} cos \phi_{4 2}
  + 0.26 \mu_0^{\prime} M_{30} cos \phi_{4 3},
\,\end{eqnarray}
\subsection{High $tan\beta$ regime}
\begin{eqnarray}
m_{H_u}^2(t_Z)&=& 0.014  A_{b_0}^2 - 0.0012  A_{b_0} A_{\tau_0} cos \phi_{b \tau} -
0.0017  A_{b_0} M_{10} cos \phi_{b 1} \nonumber\\ &-& 0.014  A_{b_0} M_{20} cos \phi_{b 2}
  - 0.065  A_{b_0} M_{30} cos \phi_{b 3}
  - 0.18  A_{b^\prime_0}^2 \nonumber\\ &+&   0.044  A_{b^\prime_0} A_{\tau^\prime_0} cos \phi_{b^\prime \tau^\prime}
  -  0.035  A_{b^\prime_0}\mu_0^{\prime} cos \phi_{b^\prime 4}
  - 0.083  A_{t_0}^2 \nonumber\\ &+&  0.01  A_{t_0} A_{b_0} cos \phi_{t b}
  -  0.00053  A_{t_0} A_{\tau_0} cos \phi_{t \tau}
  + 0.01  A_{t_0} M_{10} cos \phi_{t 1} \nonumber\\ &+&  0.06  A_{t_0} M_{20} cos \phi_{t 2}
  +  0.27  A_{t_0} M_{30} cos \phi_{t 3}
  - 0.0011  A_{\tau_0}^2 \nonumber\\ &+&  0.00028  A_{\tau_0} M_{10} cos \phi_{\tau 1}
  +  0.0014  A_{\tau_0} M_{20} cos \phi_{\tau 2}
  + 0.0049  A_{\tau_0} M_{30} cos \phi_{\tau 3} \nonumber\\ &-& 0.056  A_{\tau^\prime_0}^2
  -  0.031  A_{\tau^\prime_0}\mu_0^{\prime} cos \phi_{\tau^\prime,4}
  + 0.096  A_{t^\prime_0}^2 \nonumber\\ &-& 0.032  A_{t^\prime_0} A_{b^\prime_0} cos \phi_{t^\prime b^\prime}
  +  0.0049  A_{t^\prime_0} A_{\tau^\prime_0} cos \phi_{t^\prime \tau^\prime}
  +  0.03  A_{t^\prime_0}\mu_0^{\prime} cos \phi_{t^\prime 4} \nonumber\\ &+&  0.0013  M_{10}^2
  - 0.005  M_{10} M_{20} cos \phi_{1 2}
  -  0.025  M_{10} M_{30} cos \phi_{1 3} \nonumber\\ &+&  0.2  M_{20}^2
  - 0.17  M_{20} M_{30} cos \phi_{2 3}
  - 2.6  M_{30}^2 \nonumber\\ &+&   0.029  m_{{b_R}0}^2
  + 0.028  m_{{H_d}0}^2
  + 0.6  m_{{H_u}0}^2 \nonumber\\ &-& 0.0016  m_{{l_L}0}^2
  - 0.0016  m_{{l_R}0}^2
  -  0.37  m_{{t_L}0}^2 \nonumber\\ &-& 0.4  m_{{t_R}0}^2
  - 0.0083 \mu_0^{\prime 2},
\,\end{eqnarray}
\begin{eqnarray}
m_{H_d}^2(t_Z)&=& -0.11  A_{b_0}^2 + 0.033  A_{b_0} A_{\tau_0} cos \phi_{b \tau} + 0.0025  A_{b_0}
M_{10} cos \phi_{b 1} \nonumber\\ &+&   0.069  A_{b_0} M_{20} cos \phi_{b 2}
  + 0.36  A_{b_0} M_{30} cos \phi_{b 3}
  + 0.051  A_{b^\prime_0}^2 \nonumber\\ &-& 0.0074  A_{b^\prime_0} A_{\tau^\prime_0} cos \phi_{b^\prime \tau^\prime}
  -  0.00043  A_{b^\prime_0}\mu_0^{\prime} cos \phi_{b^\prime 4}
   + 0.009  A_{t_0}^2 \nonumber\\ &+&  0.021  A_{t_0} A_{b_0} cos \phi_{t b}
  -  0.0032  A_{t_0} A_{\tau_0} cos \phi_{t \tau}
  - 0.0013  A_{t_0} M_{10} cos \phi_{t 1} \nonumber\\ &-& 0.014  A_{t_0} M_{20} cos \phi_{t 2}
  -  0.069  A_{t_0} M_{30} cos \phi_{t 3}
  - 0.046  A_{\tau_0}^2 \nonumber\\ &+&  0.0096  A_{\tau_0} M_{10} cos \phi_{\tau 1}
  +  0.018  A_{\tau_0} M_{20} cos \phi_{\tau 2}
  - 0.053  A_{\tau_0} M_{30} cos \phi_{\tau 3} \nonumber\\ &+&  0.011  A_{\tau^\prime_0}^2
  +  0.0045  A_{\tau^\prime_0}\mu_0^{\prime} cos \phi_{\tau^\prime 4}
  - 0.24  A_{t^\prime_0}^2 \nonumber\\ &-& 0.025  A_{t^\prime_0} A_{b^\prime_0} cos \phi_{t^\prime b^\prime}
  +  0.001  A_{t^\prime_0} A_{\tau^\prime_0} cos \phi_{t^\prime \tau^\prime}
  -  0.11  A_{t^\prime_0}\mu_0^{\prime} cos \phi_{t^\prime 4} \nonumber\\ &+&  0.011  M_{10}^2
  - 0.005  M_{10} M_{20} cos \phi_{1 2}
  -  0.0055  M_{10} M_{30} cos \phi_{1 3} \nonumber\\
  &+&  0.22  M_{20}^2
  - 0.16  M_{20} M_{30} cos \phi_{2 3}
  - 2.1  M_{30}^2 \nonumber\\
  &-& 0.31  m_{{b_R}0}^2
  + 0.61  m_{{H_d}0}^2
  + 0.03  m_{{H_u}0}^2 \nonumber\\
  &-& 0.077  m_{{l_L}0}^2
  - 0.077  m_{{l_R}0}^2
  - 0.28  m_{{t_L}0}^2 \nonumber\\
  &+&   0.03  m_{{t_R}0}^2
  + 0.13 \mu_0^{\prime 2},
\,\end{eqnarray}
\begin{eqnarray}
m_{t_L}^2(t_Z)&=& -0.036  A_{b_0}^2 + 0.004  A_{b_0} A_{\tau_0} cos \phi_{b \tau} + 0.0013  A_{b_0}
M_{10} cos \phi_{b 1} \nonumber\\ &+&   0.022  A_{b_0} M_{20} cos \phi_{b 2}
  + 0.1  A_{b_0} M_{30} cos \phi_{b 3}
  - 0.041  A_{b^\prime_0}^2 \nonumber\\ &+&   0.0048  A_{b^\prime_0} A_{\tau^\prime_0} cos \phi_{b^\prime \tau^\prime}
  -  0.015  A_{b^\prime_0}\mu_0^{\prime} cos \phi_{b^\prime 4}
  - 0.024  A_{t_0}^2 \nonumber\\ &+&  0.011  A_{t_0} A_{b_0} cos \phi_{t b}
  -  0.00083  A_{t_0} A_{\tau_0} cos \phi_{t \tau}
  + 0.0028  A_{t_0} M_{10} cos \phi_{t 1} \nonumber\\ &+&  0.015  A_{t_0} M_{20} cos \phi_{t 2}
  +  0.067  A_{t_0} M_{30} cos \phi_{t 3}
  + 0.0046  A_{\tau_0}^2 \nonumber\\ &-& 0.00095  A_{\tau_0} M_{10} cos \phi_{\tau 1}
  -  0.0042  A_{\tau_0} M_{20} cos \phi_{\tau 2}
  - 0.011  A_{\tau_0} M_{30} cos \phi_{\tau 3} \nonumber\\ &+&  0.0065  A_{\tau^\prime_0}^2
  +  0.0046  A_{\tau^\prime_0}\mu_0^{\prime} cos \phi_{\tau^\prime 4}
  - 0.052  A_{t^\prime_0}^2 \nonumber\\ &-& 0.019  A_{t^\prime_0} A_{b^\prime_0} cos \phi_{t^\prime b^\prime}
  +  0.0014  A_{t^\prime_0} A_{\tau^\prime_0} cos \phi_{t^\prime \tau^\prime}
   -  0.029  A_{t^\prime_0}\mu_0^{\prime} cos \phi_{t^\prime 4} \nonumber\\ &-& 0.011  M_{10}^2
   - 0.0019  M_{10} M_{20} cos \phi_{1 2}
   -  0.011  M_{10} M_{30} cos \phi_{1 3} \nonumber\\ &+&  0.32  M_{20}^2
  - 0.11  M_{20} M_{30} cos \phi_{2 3}
  + 4.7  M_{30}^2 \nonumber\\ &-& 0.098  m_{{b_R}0}^2
  - 0.091  m_{{H_d}0}^2
  - 0.12  m_{{H_u}0}^2 \nonumber\\ &+&  0.0072  m_{{l_L}0}^2
  + 0.0072  m_{{l_R}0}^2
  +  0.78  m_{{t_L}0}^2 \nonumber\\ &-& 0.12  m_{{t_R}0}^2
  + 0.35 \mu_0^{\prime 2},
\,\end{eqnarray}
\begin{eqnarray}
m_{t_R}^2(t_Z)&=& 0.0094  A_{b_0}^2 - 0.00082  A_{b_0} A_{\tau_0} cos \phi_{b \tau} - 0.0011  A_{b_0}
M_{10} cos \phi_{b 1} \nonumber\\ &-& 0.0095  A_{b_0} M_{20} cos \phi_{b 2}
  - 0.043  A_{b_0} M_{30} cos \phi_{b 3}
  + 0.064  A_{b^\prime_0}^2 \nonumber\\ &-& 0.01  A_{b^\prime_0} A_{\tau^\prime_0} cos \phi_{b^\prime \tau^\prime}
  +  0.0051  A_{b^\prime_0}\mu_0^{\prime} cos \phi_{b^\prime 4}
  - 0.055  A_{t_0}^2 \nonumber\\ &+&  0.0067  A_{t_0} A_{b_0} cos \phi_{t b}
  -  0.00035  A_{t_0} A_{\tau_0} cos \phi_{t \tau}
  + 0.0066  A_{t_0} M_{10} cos \phi_{t 1} \nonumber\\ &+&  0.04  A_{t_0} M_{20} cos \phi_{t 2}
  +  0.18  A_{t_0} M_{30} cos \phi_{t 3}
  - 0.00076  A_{\tau_0}^2 \nonumber\\ &+&  0.00019  A_{\tau_0} M_{10} cos \phi_{\tau 1}
  +  0.00094  A_{\tau_0} M_{20} cos \phi_{\tau 2}
  + 0.0033  A_{\tau_0} M_{30} cos \phi_{\tau 3} \nonumber\\ &+&  0.017  A_{\tau^\prime_0}^2
   +  0.0073  A_{\tau^\prime_0}\mu_0^{\prime} cos \phi_{\tau^\prime 4}
  - 0.17  A_{t^\prime_0}^2 \nonumber\\ &-& 0.013  A_{t^\prime_0} A_{b^\prime_0} cos \phi_{t^\prime b^\prime}
  +  0.00024  A_{t^\prime_0} A_{\tau^\prime_0} cos \phi_{t^\prime \tau^\prime}
  -  0.078  A_{t^\prime_0}\mu_0^{\prime} cos \phi_{t^\prime 4} \nonumber\\ &+&  0.043  M_{10}^2
  - 0.0033  M_{10} M_{20} cos \phi_{1 2}
  -  0.017  M_{10} M_{30} cos \phi_{1 3} \nonumber\\ &-& 0.19  M_{20}^2
  - 0.11  M_{20} M_{30} cos \phi_{2 3}
  + 4.6  M_{30}^2 \nonumber\\ &+&   0.02  m_{{b_R}0}^2
  + 0.018  m_{{H_d}0}^2
  - 0.27  m_{{H_u}0}^2 \nonumber\\ &-& 0.0011  m_{{l_L}0}^2
  - 0.0011  m_{{l_R}0}^2
  -  0.25  m_{{t_L}0}^2 \nonumber\\ &+&  0.73  m_{{t_R}0}^2
  + 0.43 \mu_0^{\prime 2}
\,\end{eqnarray}
\begin{eqnarray}
m_{b_R}^2(t_Z)&=& -0.081  A_{b_0}^2 + 0.0089  A_{b_0} A_{\tau_0} cos \phi_{b \tau} + 0.0038  A_{b_0}
M_{10} cos \phi_{b 1} \nonumber\\ &+&   0.053  A_{b_0} M_{20} cos \phi_{b 2}
  + 0.25  A_{b_0} M_{30} cos \phi_{b 3}
  - 0.15  A_{b^\prime_0}^2 \nonumber\\ &+&   0.02  A_{b^\prime_0} A_{\tau^\prime_0} cos \phi_{b^\prime \tau^\prime}
  -  0.036  A_{b^\prime_0}\mu_0^{\prime} cos \phi_{b^\prime 4}
  + 0.0064  A_{t_0}^2 \nonumber\\ &+&  0.015  A_{t_0} A_{b_0} cos \phi_{t b}
  -  0.0013  A_{t_0} A_{\tau_0} cos \phi_{t \tau}
  - 0.0011  A_{t_0} M_{10} cos \phi_{t 1} \nonumber\\ &-& 0.01  A_{t_0} M_{20} cos \phi_{t 2}
  -  0.047  A_{t_0} M_{30} cos \phi_{t 3}
  + 0.01  A_{\tau_0}^2 \nonumber\\ &-& 0.0021  A_{\tau_0} M_{10} cos \phi_{\tau 1}
  -  0.0093  A_{\tau_0} M_{20} cos \phi_{\tau 2}
  - 0.025  A_{\tau_0} M_{30} cos \phi_{\tau 3} \nonumber\\ &-& 0.0037  A_{\tau^\prime_0}^2
  +  0.0019  A_{\tau^\prime_0}\mu_0^{\prime} cos \phi_{\tau^\prime 4}
   + 0.066  A_{t^\prime_0}^2 \nonumber\\ &-& 0.025  A_{t^\prime_0} A_{b^\prime_0} cos \phi_{t^\prime b^\prime}
  +  0.0026  A_{t^\prime_0} A_{\tau^\prime_0} cos \phi_{t^\prime \tau^\prime}
  +  0.02  A_{t^\prime_0}\mu_0^{\prime} cos \phi_{t^\prime 4} \nonumber\\ &+&  0.01  M_{10}^2
  - 0.00056  M_{10} M_{20} cos \phi_{1 2}
  -  0.0055  M_{10} M_{30} cos \phi_{1 3} \nonumber\\ &-& 0.14  M_{20}^2
  - 0.11  M_{20} M_{30} cos \phi_{2 3}
  + 4.9  M_{30}^2 \nonumber\\ &+&   0.78  m_{{b_R}0}^2
  - 0.2  m_{{H_d}0}^2
  + 0.021  m_{{H_u}0}^2 \nonumber\\ &+&  0.015  m_{{l_L}0}^2
  + 0.015  m_{{l_R}0}^2
  - 0.2  m_{{t_L}0}^2 \nonumber\\ &+&   0.021  m_{{t_R}0}^2
  + 0.28 \mu_0^{\prime 2},
\,\end{eqnarray}
\begin{eqnarray}
m_{l_L}^2(t_Z)&=& 0.007  A_{b_0}^2 + 0.02  A_{b_0} A_{\tau_0} cos \phi_{b \tau} - 0.0032 A_{b_0} M_{10}
cos \phi_{b 1} \nonumber\\ &-& 0.011  A_{b_0} M_{20} cos \phi_{b 2}
  - 0.0082  A_{b_0} M_{30} cos \phi_{b 3}
  - 0.0049  A_{b^\prime_0}^2 \nonumber\\ &+&   0.023  A_{b^\prime_0} A_{\tau^\prime_0} cos \phi_{b^\prime \tau^\prime}
   +  0.01  A_{b^\prime_0}\mu_0^{\prime} cos \phi_{b^\prime 4}
  - 0.0005  A_{t_0}^2 \nonumber\\ &-& 0.00072  A_{t_0} A_{b_0} cos \phi_{t b}
  -  0.0013  A_{t_0} A_{\tau_0} cos \phi_{t \tau}
  + 0.00027  A_{t_0} M_{10} cos \phi_{t 1} \nonumber\\ &+&   0.0011  A_{t_0} M_{20} cos \phi_{t 2}
  + 0.0015  A_{t_0} M_{30} cos \phi_{t 3}
  - 0.062  A_{\tau_0}^2 \nonumber\\ &+&   0.013  A_{\tau_0} M_{10} cos \phi_{\tau 1}
  + 0.032  A_{\tau_0} M_{20} cos \phi_{\tau 2}
  -  0.015  A_{\tau_0} M_{30} cos \phi_{\tau 3} \nonumber\\ &-& 0.064  A_{\tau^\prime_0}^2
  -  0.04  A_{\tau^\prime_0}\mu_0^{\prime} cos \phi_{\tau^\prime 4}
  + 0.018  A_{t^\prime_0}^2 \nonumber\\ &+&   0.00067  A_{t^\prime_0} A_{b^\prime_0} cos \phi_{t^\prime b^\prime}
   +  0.0016  A_{t^\prime_0} A_{\tau^\prime_0} cos \phi_{t^\prime \tau^\prime}
  +  0.0065  A_{t^\prime_0}\mu_0^{\prime} cos \phi_{t^\prime 4} \nonumber\\ &+&  0.021  M_{10}^2
  - 0.0042  M_{10} M_{20} cos \phi_{1 2}
  +  0.0028  M_{10} M_{30} cos \phi_{1 3} \nonumber\\ &+&  0.43  M_{20}^2
  + 0.011  M_{20} M_{30} cos \phi_{2 3}
  + 0.043  M_{30}^2 \nonumber\\ &+&   0.017  m_{{b_R}0}^2
  - 0.084  m_{{H_d}0}^2
  - 0.0011  m_{{H_u}0}^2 \nonumber\\ &+&  0.9  m_{{l_L}0}^2
  - 0.1  m_{{l_R}0}^2
  +  0.016  m_{{t_L}0}^2 \nonumber\\ &-& 0.0011  m_{{t_R}0}^2
  + 0.14 \mu_0^{\prime 2},
\,\end{eqnarray}
\begin{eqnarray}
m_{l_R}^2(t_Z)&=&0.014  A_{b_0}^2 + 0.039  A_{b_0} A_{\tau_0} cos \phi_{b \tau} - 0.0063  A_{b_0}
M_{10} cos \phi_{b 1} \nonumber\\ &-& 0.023  A_{b_0} M_{20} cos \phi_{b 2}
  - 0.016  A_{b_0} M_{30} cos \phi_{b 3}
  - 0.0097  A_{b^\prime_0}^2 \nonumber\\ &+&   0.045  A_{b^\prime_0} A_{\tau^\prime_0} cos \phi_{b^\prime \tau^\prime}
  +  0.021  A_{b^\prime_0}\mu_0^{\prime} cos \phi_{b^\prime 4}
  - 0.001  A_{t_0}^2 \nonumber\\ &-& 0.0014  A_{t_0} A_{b_0} cos \phi_{t b}
  -  0.0025  A_{t_0} A_{\tau_0} cos \phi_{t \tau}
  + 0.00053  A_{t_0} M_{10} cos \phi_{t 1} \nonumber\\ &+&   0.0022  A_{t_0} M_{20} cos \phi_{t 2}
  + 0.0029  A_{t_0} M_{30} cos \phi_{t 3}
  - 0.12  A_{\tau_0}^2 \nonumber\\ &+&   0.025  A_{\tau_0} M_{10} cos \phi_{\tau 1}
  + 0.064  A_{\tau_0} M_{20} cos \phi_{\tau 2}
  -  0.03  A_{\tau_0} M_{30} cos \phi_{\tau 3} \nonumber\\ &-& 0.13  A_{\tau^\prime_0}^2
  -  0.081  A_{\tau^\prime_0}\mu_0^{\prime} cos \phi_{\tau^\prime 4}
  + 0.035  A_{t^\prime_0}^2 \nonumber\\ &+&   0.0013  A_{t^\prime_0} A_{b^\prime_0} cos \phi_{t^\prime b^\prime}
   +  0.0032  A_{t^\prime_0} A_{\tau^\prime_0} cos \phi_{t^\prime \tau^\prime}
  +  0.013  A_{t^\prime_0}\mu_0^{\prime} cos \phi_{t^\prime 4} \nonumber\\ &+&  0.12  M_{10}^2
  - 0.0083  M_{10} M_{20} cos \phi_{1 2}
  +  0.0056  M_{10} M_{30} cos \phi_{1 3} \nonumber\\ &-& 0.11  M_{20}^2
  + 0.023  M_{20} M_{30} cos \phi_{2 3}
  + 0.08  M_{30}^2 \nonumber\\ &+&   0.034  m_{{b_R}0}^2
  - 0.17  m_{{H_d}0}^2
  - 0.0022  m_{{H_u}0}^2 \nonumber\\ &-& 0.2  m_{{l_L}0}^2
  + 0.8  m_{{l_R}0}^2
  +  0.031  m_{{t_L}0}^2 \nonumber\\ &-& 0.0022  m_{{t_R}0}^2
  + 0.27 \mu_0^{\prime 2},
\,\end{eqnarray}
\begin{eqnarray}
m_{3}^2(t_Z)&=&0.0052  A_{b^\prime_0} A_{t_0} cos \phi_{b^\prime t}
 + 0.024  A_{b^\prime_0} A_{\tau_0} cos \phi_{b^\prime \tau}
+  0.0035  A_{b^\prime_0} M_{10} cos \phi_{b^\prime 1} \nonumber\\ &+&  0.062  A_{b^\prime_0} M_{20} cos
\phi_{b^\prime 2}
  +  0.31  A_{b^\prime_0} M_{30} cos \phi_{b^\prime 3}
  + 0.022  A_{\tau^\prime_0} A_{b_0} cos \phi_{\tau^\prime b} \nonumber\\ &-& 0.0016  A_{\tau^\prime_0} A_{t_0} cos \phi_{\tau^\prime t}
  - 0.062  A_{\tau^\prime_0} A_{\tau_0} cos \phi_{\tau^\prime \tau}
  +  0.0058  A_{\tau^\prime_0} M_{10} cos \phi_{\tau^\prime 1} \nonumber\\ &+&  0.0098  A_{\tau^\prime_0} M_{20} cos \phi_{\tau^\prime 2}
  -  0.037  A_{\tau^\prime_0} M_{30} cos \phi_{\tau^\prime 3}
  - 0.0057  A_{t^\prime_0} A_{b_0} cos \phi_{t^\prime b} \nonumber\\ &-& 0.21  A_{t^\prime_0} A_{t_0} cos \phi_{t^\prime t}
  + 0.0019  A_{t^\prime_0} A_{\tau_0} cos \phi_{t^\prime \tau}
  +  0.012  A_{t^\prime_0} M_{10} cos \phi_{t^\prime 1} \nonumber\\ &+&  0.078  A_{t^\prime_0} M_{20} cos \phi_{t^\prime 2}
  +  0.35  A_{t^\prime_0} M_{30} cos \phi_{t^\prime 3}
  - 0.036  M_{10}^2 \nonumber\\ &-& 0.36  M_{20}^2
  + 0.68  m_{30}^2
  -  0.015 \mu_0^{\prime} A_{b_0} cos \phi_{4 b} \nonumber\\ &-& 0.042 \mu_0^{\prime} A_{t_0} cos \phi_{4 t}
  -  0.017 \mu_0^{\prime} A_{\tau_0} cos \phi_{4 \tau}
   + 0.0065 \mu_0^{\prime} M_{10} cos \phi_{4 1} \nonumber\\ &+&   0.037 \mu_0^{\prime} M_{20} cos \phi_{4 2}
  + 0.14 \mu_0^{\prime} M_{30} cos \phi_{4 3}.
\,\end{eqnarray}
\subsection{trilinear terms in the NHSSM}
\label{apptriliners}
At the low values of  $tan\beta$:
\begin{eqnarray} A_t(t_Z)&=&-0.00063 A_{b_0} + 0.22 A_{t_0} + 3.6 \times 10^{-7}   A_{\tau_0} - 0.029 M_{10} - 0.23
M_{20} - 1.9M_{30} \nonumber\\
A_b(t_Z)&=&0.99 A_{b_0} - 0.13 A_{t_0} - 0.00079 A_{\tau_0} - 0.033 M_{10} - 0.48 M_{20} - 3 M_{30} \nonumber\\
A_\tau(t_Z)&=&-0.0032 A_{b_0} + 0.00029 A_{t_0} +  A_{\tau_0} - 0.16 M_{10} - 0.53 M_{20} + 0.005 M_{30}\nonumber\\
A_{t^\prime}(t_Z)&=&0.00061 A_{b^\prime_0} - 2.8 \times 10^{-7}   A_{\tau^\prime_0} + 0.49 A_{t^\prime_0} +
0.46 \mu_0^{\prime} \nonumber\\
A_{b^\prime}(t_Z)&=&0.63 A_{b^\prime_0} - 0.00044 A_{\tau^\prime_0} + 0.14 A_{t^\prime_0} + 0.19 \mu_0^{\prime}
\nonumber\\
A_{\tau^\prime}(t_Z)&=& -0.0018 A_{b^\prime_0} + 0.49 A_{\tau^\prime_0} - 0.00026 A_{t^\prime_0} + 0.47
\mu_0^{\prime}. \,\end{eqnarray}
When  $tan\beta$ is high:
 \begin{eqnarray} A_t(t_Z)&=& -0.05 A_{b_0} + 0.21 A_{t_0} + 0.0045 A_{\tau_0} - 0.027 M_{10} -
0.21 M_{20} - 1.8 M_{30}  \nonumber\\
A_b(t_Z)&=& 0.38 A_{b_0} - 0.072 A_{t_0} - 0.055 A_{\tau_0} - 0.0092 M_{10} - 0.25 M_{20} - 2.1 M_{30} \nonumber\\
A_\tau(t_Z)&=& -0.26 A_{b_0} + 0.027 A_{t_0} + 0.62 A_{\tau_0} - 0.11 M_{10} - 0.32
M_{20} + 0.44 M_{30}  \nonumber\\
A_{t^\prime}(t_Z)&=& 0.082 A_{b^\prime_0} - 0.0065 A_{\tau^\prime_0} + 0.42 A_{t^\prime_0} + 0.18 \mu_0^{\prime}
 \nonumber\\
A_{b^\prime}(t_Z)&=& 0.54 A_{b^\prime_0} - 0.069 A_{\tau^\prime_0} + 0.12A_{t^\prime_0} + 0.083 \mu_0^{\prime}  \nonumber\\
A_{\tau^\prime}(t_Z)&=& -0.29 A_{b^\prime_0} + 0.6 A_{\tau^\prime_0} - 0.036 A_{t^\prime_0} + 0.4
\mu_0^{\prime}.\,\end{eqnarray}
Note that for the same values of  $tan\beta$ one-loop
MSSM results can be obtained from the NHSSM solutions via the appropriate transformations
(see text for details).
\section{MSSM $\&$ NHSSM under universality assumption}
\label{appuni}
For the sake of simplicity and completeness, we also provide the  solutions using (\ref{mintrans}), both in the
MSSM and NHSSM; $mass^2$ and trilinear terms are presented in the following subsections.
\subsection{MSSM under universal terms}
With the help of  (\ref{mintrans}) for low $tan\beta$ MSSM results are
\begin{eqnarray}
m_{H_u}^2(t_Z)&=&-0.087 A_0^2  + 0.38 A_0 M - 2.8 M^2  - 0.16 m_0^2,\nonumber \\
m_{H_d}^2(t_Z)&=&-0.0033 A_0^2  + 0.011 A_0 M + 0.49 M^2  + 0.99 m_0^2,\nonumber \\
m_{t_L}^2(t_Z)&=&-0.03 A_0^2  + 0.13 A_0 M + 5.7 M^2  + 0.61 m_0^2,\nonumber \\
m_{t_R}^2(t_Z)&=&-0.058 A_0^2  + 0.25 A_0 M + 4.1 M^2  + 0.22 m_0^2,\nonumber \\
m_{b_R}^2(t_Z)&=&-0.0017 A_0^2  + 0.0072 A_0 M + 6.3 M^2  + 0.99 m_0^2,\nonumber \\
m_{l_L}^2(t_Z)&=&-0.00078 A_0^2  + 0.00067 A_0 M + 0.52 M^2  +  m_0^2,\nonumber \\
m_{l_R}^2(t_Z)&=&-0.0016 A_0^2  + 0.0013 A_0 M + 0.15 M^2  +  m_0^2,\nonumber \\
m_{3}^2(t_Z)&=&-0.38 A_0 \mu_0 + 0.96 m_{30}^2 + 0.26 M \mu_0,\nonumber \\
A_{t}(t_Z)&=&0.22 A_0 - 2.2 M,\nonumber \\
A_{b}(t_Z)&=&0.074 A_0  - 0.3 M,\nonumber \\
A_{\tau}(t_Z)&=&0.052 A_0  - 0.036 M,
\end{eqnarray}
for high $tan\beta$ MSSM results can be written as
\begin{eqnarray}
m_{H_u}^2(t_Z)&=&-0.061 A_0^2  + 0.27 A_0 M - 2.6 M^2  - 0.12 m_0^2,\nonumber \\
m_{H_d}^2(t_Z)&=&-0.1 A_0^2  + 0.32 A_0 M - 2. M^2  - 0.066 m_0^2,\nonumber \\
m_{t_L}^2(t_Z)&=&-0.041 A_0^2  + 0.19 A_0 M + 4.9 M^2  + 0.36 m_0^2,\nonumber \\
m_{t_R}^2(t_Z)&=&-0.041 A_0^2  + 0.18 A_0 M + 4.3 M^2  + 0.25 m_0^2,\nonumber \\
m_{b_R}^2(t_Z)&=&-0.042 A_0^2  + 0.21 A_0 M + 4.7 M^2  + 0.46 m_0^2,\nonumber \\
m_{l_L}^2(t_Z)&=&-0.037 A_0^2  + 0.0099 A_0 M + 0.51 M^2  + 0.75 m_0^2,\nonumber \\
m_{l_R}^2(t_Z)&=&-0.075 A_0^2  + 0.02 A_0 M + 0.12 M^2  + 0.49 m_0^2,\nonumber \\
m_{3}^2(t_Z)&=&-0.5 A_0 \mu_0 + 0.68 m_{30}^2 + 0.59 M \mu_0,\nonumber \\
A_{t}(t_Z)&=&0.16 A_0  - 2 M,\nonumber \\
A_{b}(t_Z)&=&0.21 A_0 - 2 M,\nonumber \\
A_{\tau}(t_Z)&=&0.2 A_0 + 0.0041 M.
\end{eqnarray}
\subsection{NHSSM under universal terms}
with the help of  (\ref{mintrans}) again for low $tan\beta$ $mass^2$ terms:
\begin{eqnarray}
m_{H_u}^2(t_Z)&=& -0.087 A_0^2 + 0.10 A_0^{\prime 2} - 0.16 m_0^2 - 2.84 M^2 + 0.067 A_0^\prime
\mu_0^{\prime} + 0.14 \mu_0^{\prime 2} +  0.38 A_0 M,  \nonumber \\
m_{H_d}^2(t_Z)&=& -0.0033 A_0^2 - 0.37 A_0^{\prime 2} + 0.99 m_0^2 + 0.49 M^2 - 0.31 A_0^\prime
\mu_0^{\prime} + 0.6 \mu_0^{\prime 2} + 0.011 A_0 M,  \nonumber \\
m_{t_L}^2(t_Z)&=& -0.03 A_0^2 - 0.089 A_0^{\prime 2} + 0.61 m_0^2 + 5.7 M^2 - 0.08 A_0^\prime \mu_0^{\prime} +
0.3 \mu_0^{\prime 2} + 0.13 A_0 M,  \nonumber \\
m_{t_R}^2(t_Z)&=& -0.058 A_0^2 - 0.18 A_0^{\prime 2} + 0.22 m_0^2 + 4.1 M^2 - 0.16 A_0^\prime \mu_0^{\prime} +
0.6 \mu_0^{\prime 2} + 0.25 A_0 M,  \nonumber \\
m_{b_R}^2(t_Z)&=& -0.0017 A_0^2 - 0.00079 A_0^{\prime 2} + 0.99 m_0^2 + 6.3 M^2 - 0.00015 A_0^\prime
\mu_0^{\prime} + 0.0029 \mu_0^{\prime 2} \nonumber \\ &+&
 0.0072 A_0 M,  \nonumber \\
m_{l_L}^2(t_Z)&=& -0.00078 A_0^2 - 0.00023 A_0^{\prime 2} +  m_0^2 + 0.52 M^2 - 0.00022 A_0^\prime
\mu_0^{\prime} + 0.0012 \mu_0^{\prime 2} \nonumber \\
&+&  0.00067 A_0 M,  \nonumber \\
m_{l_R}^2(t_Z)&=& -0.0016 A_0^2 - 0.00045 A_0^{\prime 2} + m_0^2 + 0.15 M^2 - 0.00044 A_0^\prime
\mu_0^{\prime} + 0.0025 \mu_0^{\prime 2} \nonumber \\ &+&  0.0013 A_0 M,  \nonumber \\
m_{3}^2(t_Z)&=& -0.27 A_0 A_0^\prime - 0.56 M^2 + 0.96  m_{30}^2 - 0.099 A_0 \mu_0^{\prime} + 0.5 A_0^\prime M +
0.32 \mu_0^{\prime} M,\, \nonumber \\
 A_t(t_Z)&=& 0.22 A_0 - 2.2 M, \nonumber \\
A_b(t_Z)&=& 0.86 A_0 - 3.6 M, \nonumber \\
A_\tau(t_Z)&=& 0.99 A_0 - 0.68 M, \nonumber \\
A_{t^\prime}(t_Z)&=& 0.49 A_0^\prime + 0.46 \mu_0^{\prime}, \nonumber \\
A_{b^\prime}(t_Z)&=&0.77 A_0^\prime + 0.19 \mu_0^{\prime}, \nonumber \\
A_{\tau^\prime}(t_Z)&=& 0.49 A_0^\prime + 0.47 \mu_0^{\prime}.
\end{eqnarray}
For high $tan\beta$:
\begin{eqnarray}
m_{H_u}^2(t_Z)&=& -0.061 A_0^2 - 0.12 A_0^{\prime 2} - 0.12 m_0^2 - 2.6 M^2 - 0.036 A_0^\prime \mu_0^{\prime} -
0.0083 \mu_0^{\prime 2} +  0.27 A_0 M,  \nonumber \\
m_{H_d}^2(t_Z)&=& -0.1 A_0^2 - 0.21 A_0^{\prime 2} - 0.066 m_0^2 - 2. M^2 - 0.11 A_0^\prime
\mu_0^{\prime} + 0.13 \mu_0^{\prime 2} + 0.32 A_0 M,  \nonumber \\
m_{t_L}^2(t_Z)&=& -0.041 A_0^2 - 0.1 A_0^{\prime 2} + 0.36 m_0^2 + 4.9 M^2 - 0.04 A_0^\prime \mu_0^{\prime} +
0.35 \mu_0^{\prime 2} + 0.19 A_0 M,  \nonumber \\
m_{t_R}^2(t_Z)&=& -0.041 A_0^2 - 0.11 A_0^{\prime 2} + 0.25 m_0^2 + 4.3 M^2 - 0.065 A_0^\prime \mu_0^{\prime} +
0.43 \mu_0^{\prime 2} + 0.18 A_0 M,  \nonumber \\
m_{b_R}^2(t_Z)&=& -0.042 A_0^2 - 0.086 A_0^{\prime 2} + 0.46 m_0^2 + 4.7 M^2 - 0.014 A_0^\prime \mu_0^{\prime} +
0.28 \mu_0^{\prime 2} + 0.21 A_0 M,  \nonumber \\
m_{l_L}^2(t_Z)&=& -0.037 A_0^2 - 0.027 A_0^{\prime 2} + 0.75 m_0^2 + 0.51 M^2 - 0.023 A_0^\prime \mu_0^{\prime}
+ 0.14 \mu_0^{\prime 2} \nonumber \\ &+&  0.0099 A_0 M,  \nonumber \\
m_{l_R}^2(t_Z)&=& -0.075 A_0^2 - 0.054 A_0^{\prime 2} + 0.49 m_0^2 + 0.11 M^2 - 0.047 A_0^\prime \mu_0^{\prime}
+ 0.27 \mu_0^{\prime 2} +  0.02 A_0 M,  \nonumber \\
m_{3}^2(t_Z)&=& -0.23 A_0 A_0^\prime - 0.4 M^2 + 0.68  m_{30}^2 - 0.074 A_0 \mu_0^{\prime} + 0.8 A_0^\prime M +
0.19 \mu_0^{\prime} M, \nonumber \\
A_t(t_Z)&=&0.16 A_0 - 2.1 M, \nonumber \\
A_b(t_Z)&=&0.25 A_0 - 2.3 M \nonumber \\
A_\tau(t_Z)&=&0.39 A_0 + 0.0081 M, \nonumber \\
A_{t^\prime}(t_Z)&=&0.5 A_0^\prime + 0.18 \mu_0^{\prime}, \nonumber \\
A_{b^\prime}(t_Z)&=&0.6 A_0^\prime + 0.083 \mu_0^{\prime}, \nonumber \\
A_{\tau^\prime}(t_Z)&=& 0.28 A_0^\prime + 0.4 \mu_0^{\prime}.
\end{eqnarray}

\end{document}